\documentclass[prb,twocolumn,showpacs,preprintnumbers,english,amsmath,amssymb]{revtex4}

\usepackage{graphicx,epsf,here}
\usepackage{dcolumn}
\usepackage{bm}

\newcommand{\avec}[1]{\stackrel{\to }{#1}}

\begin{document}

\title{Theory of vortices in hybridized ballistic/diffusive-band superconductors}
\author{K. Tanaka,$^{1,2}$ M. Eschrig,$^{2}$ and D. F. Agterberg$^{3}$} 
\affiliation{
$^{1}$Department of Physics and Engineering Physics, University of Saskatchewan,
Saskatoon, Saskatchewan, Canada S7N 5E2 \\
$^{2}$Institut f\"ur Theoretische Festk\"orperphysik and DFG-Center for Functional Nanostructures,
Universit\"at Karlsruhe, D-76128 Karlsruhe, Germany\\
$^{3}$Department of Physics, University of Wisconsin -
Milwaukee, P.O. Box 413, Milwaukee, Wisconsin 53211, USA}
\date{11 February 2007; published in Phys. Rev. B {\bf 75}, 214512 (2007)}
\begin{abstract}
We study the electronic structure in the vicinity of a vortex in a two-band
superconductor in which the quasiparticle motion is ballistic in one band and
diffusive in the other.  
This study is based on a model appropriate for such a case,
that we have introduced recently [Tanaka {\it et al.}, Phys. Rev. B {\bf 73}, 220501(R) (2006)].
We argue that in the two-band superconductor MgB$_2$, such a case is
realised. Motivated by the experimental findings on MgB$_2$, 
we assume that superconductivity in the diffusive band is 
``weak,'' i.e., mostly induced.
We examine intriguing features of the order parameter, the current density, and
the vortex core spectrum in the ``strong'' ballistic band under the influence of 
hybridization with the ``weak'' diffusive band.  
Although the order parameter in the diffusive band is induced, the
characteristic length scales in the two bands differ due to Coulomb interactions.
The current density in the vortex core is dominated by the contribution 
from the ballistic band, while outside the core the contribution 
from the diffusive band can be substantial, or even dominating.  
The current density in the diffusive band has strong temperature dependence, 
exhibiting the Kramer-Pesch effect when hybridization is strong.
A particularly interesting feature of our model is the possibility of 
additional bound states near the gap edge in the ballistic band, that
are prominent in the vortex centre spectra. This contrasts with the 
single band case, where there is no gap-edge bound state in the vortex 
centre.
We find the above-mentioned 
unique features for parameter values relevant for MgB$_2$.
\end{abstract}

\pacs{74.20.-z, 74.50.+r, 74.70.Ad, 74.81.-g}
\maketitle

\section{Introduction}
\label{intro}

One can learn a great deal about the pairing mechanism of a superconductor from
the electronic properties in the presence of inhomogeneity, such as vortices and
impurities.
Although multiple-band superconductivity was first studied almost 50 years ago,
\cite{suhl59}
the structure of a vortex in multiband superconductors has not been well understood,
especially when impurities are present.
The best material that has been discovered so far for studying
multiband superconductivity is MgB$_2$.\cite{nagamatsu01}  The consensus is that superconductivity
is driven by electron-phonon interactions, and that it can be well
described by a two-band model, with the ``strong'' $\sigma$ band (energy gap
$\Delta_\sigma\approx 7.2$ meV) and the ``weak'' $\pi$ band 
($\Delta_\pi\approx 2.3$ meV).\cite{shulga01,liu01,bouguet01,schmidt02,iavarone02,golubov02,choi02,mazin04,
kwok03,dahm05}
The two energy gaps vanish at a common transition temperature $T_c$ (Refs.~\onlinecite{szabo01,gonnelli02,holanova04}) and there is evidence of induced superconductivity
in the $\pi$ band.\cite{schmidt02,eskildsen02,geerk05,giubileo06} 

The one unusual aspect of this material as a multiband superconductor is the effect
of impurities.  The standard theory\cite{suhl59,tang70,golubov97} 
tells us that interband scattering
by nonmagnetic impurities
should reduce $T_c$ and the gap ratio.
Mazin and co-workers,\cite{mazin02,erwin03} however, have shown theoretically that this
does not apply to MgB$_2$, due to different symmetries of the $\sigma$ and $\pi$
orbitals, and hence negligible interband scattering.
Moreover, impurities or defects, in particular those at Mg-sites which tend to occur
more easily than at B-sites, affect only the $\pi$ band strongly.
Indeed, many experiments have shown that the $\sigma$ and $\pi$ bands are essentially 
in the ballistic and diffusive limit, respectively.\cite{mazin02,erwin03,putti03,putti04,putti05,quilty03,ortolani05,carrington05,pallecchi06,samuely06}
Even in samples in which the $\sigma$ band is influenced by impurities substantially,
two-gap superconductivity is retained.\cite{holanova04,samuely06,wang03,samuely03,schmidt03,putti68,ribeiro03,angst05,tsuda05,braccini05,samuely05,klein06}
The question of whether or not the two gaps merge in dirty samples has not yet been
settled.\cite{samuely05,klein06,gonnelli04,kortus05}

Due to induced superconductivity and possibly also owing to 
these peculiar effects of impurities, 
the electronic structure around a vortex in MgB$_2$ has been found to 
exhibit intriguing properties.
Eskildsen {\it et al.},\cite{eskildsen02} by tunneling along the $c$ axis,
have probed the vortex core structure in the $\pi$ band.  The local density of states
(LDOS) was found to be completely flat as a function of energy at the vortex core,
showing no sign of bound states.  Also the `core size' ($\sim$50nm)
as measured by a decay length of the zero-bias LDOS turned out to be 
much longer than expected from $H_{c2}$ ($\sim$10nm).\cite{eskildsen02,kohen05}
Moreover, the existence of two effective coherence lengths 
in MgB$_2$ has been suggested by
the $\mu$SR measurement of a vortex lattice.\cite{serventi04}
There has been, however, no experiment directly probing the $\sigma$ band in the vortex state, and the electronic structure around a vortex in the $\sigma$ band is yet to
be determined.

Theoretically, the vortex structure in a two-band superconductor has been studied
in terms of two clean bands\cite{nakai02} and two dirty bands.\cite{koshelev03}
Neither of these models, however, applies to many MgB$_2$ samples, which have
the clean $\sigma$ and dirty $\pi$ bands.
Recently, we have formulated a unique model for a  multiband superconductor
with both a clean and a dirty band.\cite{tanaka06} 
We have studied the effects of 
induced superconductivity and impurities in the weak diffusive band on the
electronic properties  in the strong ballistic band around a vortex.
A particularly intriguing feature found in this model is the possibility of 
bound states near the gap edge in the ballistic band in the vortex core, 
in addition to the well-known Caroli-de Gennes-Matricon bound states.\cite{caroli64} Such bound states
do not exist in a single ballistic band, and they arise solely from coupling to the
diffusive band.  Our model has also been applied to study the Kramer-Pesch effect\cite{kramer74} in coupled clean and dirty bands.\cite{gumann05}
It has been found that the Kramer-Pesch effect is induced in the dirty band, 
which is absent when there is no coupling with the clean band.
Thus, hybridization of ballistic and diffusive bands can lead to unusual properties
of a multiband superconductor, and the vortex core structure is an example in which
the effects of induced superconductivity and impurities manifest clearly.

In this work, we make an extensive study of the electronic structure in the vicinity
of a vortex in a strong ballistic band, under the influence of  hybridization with a
weak diffusive band.   Our model is based on coupled Eilenberger and Usadel equations,
which are solved directly and numerically. 
We assume that interband scattering by impurities is negligible and that
the two bands are coupled only by the pairing interaction, 
as justified for typical MgB$_2$ samples.
Unique features of the
order parameter, the current density, and the vortex core spectrum 
are examined in detail.
In particular, we study the development of the gap-edge bound states as various physical
parameters are varied.
Although this work has been motivated by the impurity effects
\cite{holanova04,mazin02,erwin03,putti03,putti04,putti05,quilty03,ortolani05,carrington05,pallecchi06,samuely06,wang03,samuely03,schmidt03,putti68,ribeiro03,angst05,tsuda05,braccini05,samuely05,klein06}
and the STM measurement of the vortex state in MgB$_2$,
\cite{eskildsen02,kohen05} our model is general and applicable to any superconductor
in which ballistic and diffusive bands are coupled mainly by the 
pairing interaction.

The paper is organized as follows. The formulation and computational details are given
in Secs.~\ref{theory} and \ref{comp}, respectively.  Results are presented in 
Sec.~\ref{results} and they are summarised and discussed in Sec.~\ref{conclusion}.
Throughout the paper we use units with $\hbar = c = 1$.

\section{Theoretical description}
\label{theory}

Both the ballistic and diffusive limits of superconductivity can be described
within one unified theory, the quasiclassical theory of superconductivity.\cite{larkin68,eilenberger,usadel70,serene83,rammer86,larkin86,FLT} 
The central quantity of this theory, containing all the 
physical information, is the quasiclassical Green function, or
propagator, $\hat{g}(\epsilon, {\bf p}_{F\alpha}, {\bf R})$. Here
$\epsilon$ is the quasiparticle energy measured from the chemical
potential, ${\bf p}_{F\alpha}$ the quasiparticle momentum on the Fermi
surface of band $\alpha$, 
and ${\bf R}$ is the spatial coordinate. The hat refers
to the 2$\times$2 matrix structure of the propagator in the Nambu-Gor'kov
particle-hole space.  
In the ballistic case, the equation of motion for $\hat{g}$ is the
Eilenberger equation, and in the diffusive case the Usadel equation.
Our model is appropriate for any two-band
superconductor with a clean and a dirty band (generalisation to several
bands is straightforward). However, having in mind $\rm MgB_2$, for
definiteness we call the two bands $\sigma$ and $\pi$ bands, respectively.

In the clean $\sigma$ band, $\hat g_\sigma(\epsilon, {\bf
p}_{F\sigma}, {\bf R})$ satisfies the Eilenberger equation,
\cite{larkin68,eilenberger}
\begin{equation}
\left[ \epsilon \hat \tau_3 - \hat \Delta_\sigma 
,\; \hat g_\sigma \right]
+i {\bf v}_{F\sigma} \cdot {\nabla }  \hat g_\sigma
= \hat 0,
\label{eil}
\end{equation}
where ${\bf v}_{F\sigma}$ is the Fermi velocity and $\hat \Delta_\sigma$
the (spatially varying) order parameter.
The three Pauli matrices in Nambu-Gor'kov space are denoted by 
$\hat \tau_i$, $i=1,2,3$, and $[...,... ]$ denotes the commutator.
Throughout this work, we
ignore the variation of the magnetic field in the vortex core,
assuming a strongly type-II superconductor
(this is justified for example for MgB$_2$).

For the $\pi$ band we assume that it is in the diffusive limit. 
In the presence of strong impurity scattering, the momentum dependence of the
quasiclassical Green function is averaged out, and 
the equation of motion for the resulting propagator
$\hat g_\pi(\epsilon,{\bf R})$ reduces to the
Usadel equation,\cite{usadel70} 
\begin{equation}
\label{usadeleq}
\left[ \epsilon \hat \tau_3 - \hat \Delta_\pi ,\; \hat g_\pi \right] 
+ 
{\nabla }
\frac{\mathbb{D}}{\pi}
(\hat g_\pi {\nabla } \hat g_\pi )
= \hat 0,
\end{equation}
with 
the diffusion constant tensor $\mathbb{D}$.
Both ballistic and diffusive propagators are normalized according 
to\cite{larkin68}
\begin{equation}
\hat g_\sigma^2=\hat g_\pi^2=-\pi^2 \hat 1.
\label{norm}
\end{equation}

A two-band superconductor with a ballistic and a diffusive band can
exist only if interband scattering by impurities is weak.
We neglect in the following interband scattering by impurities, and assume that
the quasiparticles in different bands are coupled only
through the pairing interaction.
Self-consistency is achieved through the coupled gap equations for the 
spatially varying order parameters in each band,
\begin{equation}
\Delta_\alpha ({\bf R}) = \sum_{\beta} V_{\alpha\beta} N_{F\beta}
{\cal F}_\beta ({\bf R}),
\label{gapeq}
\end{equation}
where $\alpha,\beta \in \{\sigma,\pi\}$, and
$\hat\Delta_\alpha=\hat\tau_1~\rm{Re}~\Delta_\alpha-\hat\tau_2~\rm{Im}~\Delta_\alpha$.
The coupling matrix $V_{\alpha\beta}$ determines
the pairing interaction, $N_{F\beta}$ is the Fermi-surface density of
states on band $\beta$, and
\begin{eqnarray}
{\cal F}_{\sigma }({\bf R})&\equiv&
\int_{-\epsilon_c}^{\epsilon_c} {d\epsilon\over 2 \pi i}\,
\langle f_\sigma (\epsilon , {\bf p}_{F\sigma},{\bf R}) \rangle_{{\bf p}_{F\sigma}}
\, {\rm tanh}\Biggl({\epsilon\over 2 T}\Biggr),
\label{F1} \nonumber \\
{\cal F}_{\pi }({\bf R})&\equiv&
\int_{-\epsilon_c}^{\epsilon_c} {d\epsilon\over 2 \pi i}\,
f_\pi (\epsilon, {\bf R} )
\, {\rm tanh}\Biggl({\epsilon\over 2 T}\Biggr).
\label{F2}
\end{eqnarray}
Here $f_\alpha$ is the upper off-diagonal (1,2) element of the matrix
propagator $\hat g_\alpha$, and $\epsilon_c$ is a cutoff energy. 
The Fermi surface average over the $\sigma$-band is denoted by
$\langle\cdots\rangle_{{\bf p}_{F\sigma}}$.

Note that within our model the precise form of the Fermi surface in 
the $\pi$ band is not relevant, as in the diffusive limit all necessary
information is contained in the diffusion constant tensor $\mathbb{D}$. For
simplicity we have assumed in our calculation an isotropic tensor 
$\mathbb{D}_{ij}=D\delta_{ij}$. 
The diffusion constant defines the $\pi$-band
coherence length $\xi_{\pi }=\sqrt{D/2\pi T_c}$. 
For the $\sigma $-band Fermi surface we assume a cylindrical shape, as
motivated by the Fermi surface of MgB$_2$. This allows us to treat the
$\sigma $ band as quasi-two-dimensional, 
${\bf v}_{F\sigma}=v_{F\sigma } \hat{\bf p}_r$, with cylindrical coordinates
$(p_r,p_\phi , p_z)$ and the unit vector $\hat{\bf p}_r$ in direction
of ${\bf p}_r$.
We define the coherence length in the $\sigma $ band as
$\xi_{\sigma}=v_{F\sigma }/ 2\pi T_c$. It will be used as length unit
throughout this paper.

The numerical solution of 
the (nonlinear) system of Eqs. (\ref{eil})--(\ref{norm}) is greatly simplified by
using the Riccati parameterization of the 
Green functions, both for the ballistic case\cite{nagato93,schopohl,matthias0,matthias1} 
and for the diffusive case,\cite{matthias2} 
\begin{equation}
\hat g_\alpha = -\; \frac{i \pi }{1+\gamma_\alpha \tilde \gamma_\alpha }
\left(
\begin{array}{cc}
1-\gamma_\alpha \tilde \gamma_\alpha  & 2\gamma_\alpha \\
2\tilde \gamma_\alpha & \gamma_\alpha \tilde \gamma_\alpha -1
\end{array}\right),
\end{equation}
with $\tilde \gamma_\sigma (\epsilon, {\bf p}_{F\sigma}, {\bf R})=
\gamma^\ast_\sigma (-\epsilon^\ast, -{\bf p}_{F\sigma}, {\bf R})$ and
$\tilde \gamma_\pi (\epsilon, {\bf R})=
\gamma^\ast_\pi (-\epsilon^\ast, {\bf R})$.
The transport equation for 
$\gamma_\sigma (\epsilon, {\bf p}_{F\sigma}, {\bf R})$  is given 
by,\cite{nagato93,schopohl,matthias0,matthias1}
\begin{equation}
\Delta_\sigma + 
2 \epsilon \, \gamma_\sigma +
\Delta_\sigma^\ast \gamma_\sigma^2 +
i{\bf v}_{F\sigma} \nabla \gamma_{\sigma } = 0.
\end{equation}
and for $\gamma_\pi (\epsilon, {\bf R})$ it is\cite{matthias2}
\begin{equation}
\Delta_\pi + 
2 \epsilon \, \gamma_\pi +
\Delta_\pi^\ast \gamma_\pi^2  -
i D \left( \nabla^2 \gamma_\pi - \frac{2\tilde \gamma_\pi (\nabla \gamma_\pi )^2}{1+\gamma_\pi \tilde \gamma_\pi } \right) = 0.
\end{equation}

We solve Eqs.~(\ref{eil})--(\ref{F2}) self-consistently.
After self-consistency has been achieved for the order parameters, the (for
the $\sigma$ band, angle-resolved) LDOS in each band can be calculated by
\begin{eqnarray}
N_{\sigma }(\epsilon,{\bf p}_{F\sigma}, {\bf R})/N_{F\sigma}&=&-
~{\rm Im}~g_\sigma (\epsilon , {\bf p}_{F\sigma},{\bf R}) /\pi
,
\nonumber \\
N_{\pi }(\epsilon,{\bf R})/N_{F\pi}&=&-
~{\rm Im}~g_\pi (\epsilon,{\bf R})/\pi,
\label{LDOS}
\end{eqnarray}
where $g_\alpha$ is the upper diagonal (1,1) element of $\hat g_\alpha$.

The current density around the vortex has contributions from
both $\pi$ and $\sigma $ bands,
\begin{equation}
\label{eqcurr}
{\bf j}({\bf R})={\bf j}_{\sigma }({\bf R})+{\bf j}_{\pi}({\bf R}). 
\end{equation}
The corresponding expressions are 
\begin{eqnarray}
\frac{{\bf j}_{\sigma }({\bf R})}{2eN_{F\sigma }}&=&
\int_{-\infty}^{\infty} {d\epsilon\over 2\pi }\,
\langle {\bf v}_{F\sigma } {\rm Im}~g_\sigma \rangle_{{\bf p}_{F\sigma}}
\tanh\left( \frac{\epsilon}{2T}\right), \; \nonumber
\\
\frac{{\bf j}_{\pi}({\bf R})}{2eN_{F\pi}}&=&
\frac{\mathbb{D}}{\pi} \int_{-\infty}^{\infty} {d\epsilon\over 2\pi }\,
{\rm Im}~[f_\pi^{\ast }  {\nabla } f_\pi]
\tanh\left( \frac{\epsilon}{2T}\right), \;
\label{CD}
\end{eqnarray}
where $e=-|e|$ is the electron charge.

\section{Computational details}
\label{comp}

\subsection{Gap equations}
\label{comp1}

The first step is to write Eqs.~\eqref{gapeq} and \eqref{F2}
in a form that is independent of the cut-off energy $\epsilon_c $ and
the cut-off dependent interaction matrix $V_{\alpha \beta} $.
We diagonalize the interactions by
\begin{equation}
N_F \sum_{\beta =\pi,\sigma }\sqrt{n_\alpha } V_{\alpha \beta } \sqrt{n_\beta } 
{\cal Y}^{ (k) }_\beta = 
{\cal Y}^{ (k) }_\alpha \lambda^{(k)},
\end{equation}
where $N_F=N_{F\sigma } + N_{F\pi}$ and $n_\alpha = N_{F\alpha}/N_F$.
The band index runs over $\alpha = (\sigma,\pi )$, and 
thus we need two basis functions for a complete system, denoted by $k=0,1$.
We expand the order parameter and the anomalous Green functions in terms of 
the eigenvectors 
\begin{eqnarray}
\avec{{\cal Y}}^{ (k) } &= &
\left(
\begin{array}{c}
{\cal Y}^{ (k) }_\sigma  \\
{\cal Y}^{ (k) }_\pi\,. 
\end{array}
\right) 
\end{eqnarray}
The eigenvectors are orthonormal.
We also introduce the vectors
\begin{eqnarray}
\avec{\Delta } &= &
\left(
\begin{array}{c}
\sqrt{n_\sigma }\Delta_\sigma  \\
\sqrt{n_\pi }\Delta_\pi  
\end{array}
\right), \quad
\avec{{\cal F} }  = 
\left(
\begin{array}{c}
\sqrt{n_\sigma }
{\cal F}_\sigma  \\
\sqrt{n_\pi }{\cal F}_\pi 
\end{array}
\right) \; .
\end{eqnarray}
Using the expansion
\begin{equation}
\avec{\Delta }= \sum_{k=0,1} \Delta^{ (k)} \avec{{\cal Y}}^{ (k)} ,
\qquad
\avec{\cal F }= \sum_{k=0,1} {\cal F}^{ (k)} \avec{{\cal Y}}^{ (k)} 
\; ,
\end{equation}
with
$\Delta^{ (k)}= \avec{\cal Y}^{ (k)\ast }  \!\!\!\! \avec{\Delta } $ and
${\cal F}^{(k)}= \avec{\cal Y}^{ (k)\ast }  \!\!\!\! \avec{\cal F} $,
the gap equations are given by
\begin{equation}
\Delta^{(k)}({\bf R})=\lambda^{(k)}
{\cal F}^{(k)} ({\bf R})
\; .
\label{gapk}
\end{equation}
Without restriction we can choose the largest eigenvalue to be
$\lambda^{(0)}$. It can be eliminated together with the cutoff frequency
in favor of the transition
temperature between the normal and superconducting states, $T_c$.
Although both coupling constants $\lambda^{(0)}$ and $\lambda^{(1)}$ 
are cutoff dependent, 
the parameter $\Lambda $ defined by
\begin{equation}
\Lambda = \frac{\lambda^{(0)}\lambda^{(1)}}{\lambda^{(0)}-\lambda^{(1)}} 
\;
\end{equation}
is cutoff independent. We parameterize the two coupling constants by the
two cutoff independent quantities $T_c$ and $\Lambda $.

Near $T_c$, the subdominant order parameter $\Delta^{(1)}$ is of higher order in
$(T_c-T)$ than $\Delta^{(0)}$ is. 
The bulk gaps on the two Fermi surface sheets are thus determined near $T_c$
by $\Delta^{(0)}$ only, and are given by
\begin{equation}
\Delta_\sigma = \frac{{\cal Y}^{ (0) }_\sigma  }{\sqrt{n_\sigma }}\Delta^{(0)}\;, 
\qquad 
\Delta_\pi = \frac{{\cal Y}^{ (0) }_\pi  }{\sqrt{n_\pi }}\Delta^{(0)} 
\; .
\label{gap0only}
\end{equation}
We define the ratio of the bulk gaps on the two bands near $T_c$,
\begin{equation}
\rho:=\frac{\Delta_\pi }{\Delta_\sigma } = 
\frac{{\cal Y}^{ (0) }_\pi }{ {\cal Y}^{ (0) }_\sigma }\sqrt{\frac{n_\sigma }{n_\pi }}.
\end{equation}
We also introduce the notation $n:=\sqrt{n_\sigma /n_\pi }$.
Then the matrix of eigenvectors is given by
\begin{equation}
\left(
\begin{array}{cc}
{\cal Y}^{ (0) }_\sigma  &{\cal Y}^{ (1) }_\sigma \\
{\cal Y}^{ (0) }_\pi & {\cal Y}^{ (1) }_\pi 
\end{array}
\right)
= \frac{1}{\sqrt{n^2 +|\rho |^2}}
\left(
\begin{array}{cc}
n & -\rho^\ast\\
\rho & n
\end{array}
\right) \; .
\end{equation}

\subsection{Homogeneous solutions}
\label{comp2}
In the homogeneous case, at zero temperature
\begin{equation}
\lim_{T\to 0}\, {\cal F}_\alpha  = \Delta_{\alpha }\, \ln \frac{2\epsilon_c }{|\Delta_\alpha |},
\end{equation}
where $\epsilon_c$ is the usual BCS cutoff frequency of the order of
the Debye frequency. Using the BCS formula
\begin{equation}
\ln \frac{2\epsilon_c }{\pi T_c } = \frac{1}{\lambda^{(0)}} -\gamma
\end{equation}
($\gamma \approx 0.577$ is Euler's constant),
we obtain the zero-temperature gap equations
\begin{eqnarray}
\left(\gamma + \ln \frac{|\Delta_\alpha^0 |}{\pi T_c} \right) \Delta_\alpha^0 =
-\frac{1}{\Lambda } 
\sum_{\beta} 
{\cal Y}^{ (1) }_\alpha
{\cal Y}^{ (1)\ast }_\beta \sqrt{\frac{n_\beta }{n_\alpha }} \Delta_\beta^0\;,
\end{eqnarray}
where $\Delta_\alpha^0$ denotes the bulk order parameter at zero temperature
on band $\alpha$.
From this we can derive an equation for the ratio $\rho_0=\Delta_\pi^0 /\Delta_\sigma^0$
(assuming that the gap ratio $\rho$ is real),
\begin{eqnarray}
\label{gapratio}
\frac{\rho - \rho_0 }{n^2 + \rho^2} = \Lambda \frac{\rho_0 }{n^2+\rho_0 \rho } \ln | \rho_0 |\,,
\end{eqnarray}
and for the dominant zero-temperature gap
\begin{eqnarray}
\label{homDs}
\Delta_\sigma^0/T_c &=& \pi\, e^{-\gamma} e^{
-[\rho_0 \rho/(n^2 + \rho_0 \rho)]\ln |\rho_0|
}. 
\end{eqnarray}
Equation~(\ref{gapratio}) is a quadratic equation for $\rho $, 
which is readily solved analytically in terms of $\rho_0$, $n$ and $\Lambda $:
\begin{eqnarray}
\rho&=& 
\frac{2 n^2\rho_0(1+\Lambda\, \ln |\rho_0| )}{(n^2-\rho_0^2)+\sqrt{(n^2+\rho_0^2)^2-(2n\rho_0\, \Lambda
\ln|\rho_0|)^2 }}. \qquad
\label{rho}
\end{eqnarray}

\begin{figure}
\begin{minipage}{\columnwidth}
\includegraphics[width=1.00\columnwidth]{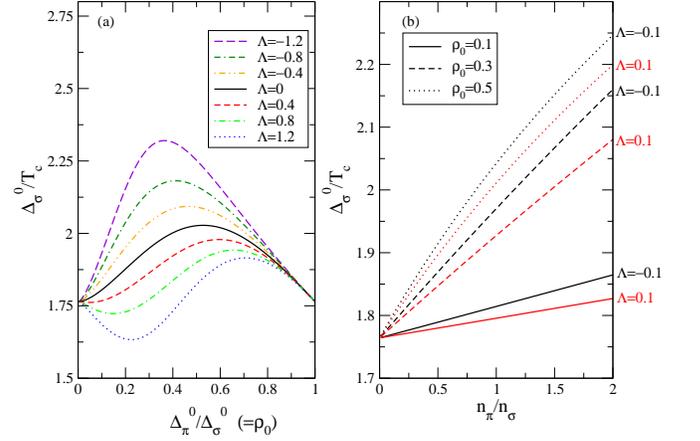}
\end{minipage}
\caption{(Color online) The bulk $|\Delta_\sigma|/T_c$ in the zero-temperature limit as given
by Eqs.~\eqref{homDs} and \eqref{rho}:
(a) as a function of $\rho_0=\Delta_\pi^0/\Delta_\sigma^0 $ for $n_{\pi}=n_{\sigma }$ and various values of $\Lambda$, and
(b) as a function of $n_{\pi}/n_{\sigma }$ for various sets of $\rho_0$ and $\Lambda$.
For a fixed  $n_{\pi }/n_{\sigma }$,
$|\Delta_\sigma|/T_c$ shows nonlinear behaviour as a function of $\rho_0$, where
the BCS value is recovered at $\rho_0=0$ and 1.
The bulk gap is enhanced as $n_{\pi }/n_{\sigma }$ increases, and this effect
is larger for larger $\rho_0$ and for smaller $\Lambda$.
}
\label{gaptc}
\end{figure}
The four material parameters $\rho_0 $, $n$, $\Lambda$, and $T_c$ completely specify
the bulk behaviour of the system.
We illustrate in Fig.~\ref{gaptc} 
how the dominant bulk gap to $T_c$ ratio at zero temperature 
is affected by $\rho_0$, $n_{\pi }/n_{\sigma }=1/n^2$, and $\Lambda$.
In (a), $\Delta_\sigma^0/T_c$ determined by
Eqs.~\eqref{homDs} and \eqref{rho} is shown as a function of $\rho_0$, for 
$n_{\pi}=n_{\sigma }$ and various values of $\Lambda$. In (b) we compare
$\Delta_\sigma^0/T_c$ as a function of 
$n_{\pi }/n_{\sigma }$ for several values of $\rho_0$, and for $\Lambda=-0.1$ and 0.1.
It can be seen in Fig.~\ref{gaptc}(a) that for a fixed $n_{\pi }/n_{\sigma }$, $\Delta_\sigma^0/T_c$ 
exhibits nonlinear behaviour as a function of $\rho_0$ depending on the value of $\Lambda$,
while the BCS value (1.764) is recovered at $\rho_0=0$ and 1.
For $\Lambda<0$ this ratio is always larger than the BCS value, and
it is always larger for $\Lambda<0$ than for  $\Lambda>0$, for fixed $\rho_0$
and $n_{\pi }/n_{\sigma }$.  When $\Lambda>0$, $\Delta_\sigma^0/T_c$ 
can be smaller than the BCS value for a certain range of $\rho_0$.
The effect becomes even more pronounced for larger $n_{\pi }/n_{\sigma }$.
For a fixed $\rho_0$, as shown in Fig.~\ref{gaptc}(b),  $\Delta_\sigma^0/T_c$ increases 
almost linearly with increasing 
$n_{\pi }/n_{\sigma }$, with larger slope for larger $\rho_0$. 
Here again, the enhancement of
$\Delta_\sigma^0/T_c$ is always stronger for $\Lambda<0$ than for  $\Lambda>0$.
For $n_{\pi }/n_{\sigma }=0$ there is only one band and the ratio reduces to the BCS value.

In fact, the enhancement of the bulk zero-temperature 
gap by stronger Coulomb interactions in the subdominant channel
can be estimated for small $|\Lambda |$ analytically 
from  Eqs.~\eqref{homDs} and \eqref{rho}, which give 
\begin{equation}
\left(\frac{\partial }{\partial \Lambda} 
\frac{\Delta_\sigma^0}{T_c}\right)_{\Lambda=0}
=-\left(\frac{\Delta_\sigma^0 }{T_c}\right)_{\Lambda=0} 
\left(\frac{n\rho_0\, \ln |\rho_0|}{n^2+\rho_0^2}\right)^2,
\label{lambda0}
\end{equation}
and $\Delta_\pi^0(\Lambda )= \rho_0 \Delta_\sigma^0(\Lambda )$.
More generally, at low $T$ for given $\rho_0$ and $n$,
the homogeneous gap over $T_c$ ratio increases with decreasing 
$\Lambda $ in both bands.
The slope at $\rho_0=1$ in Fig.~\ref{gaptc}(a) is independent of $\Lambda $
and is given by $-\pi e^{-\gamma }/(1+n^2)$. An expansion for 
small $\rho_0$ or for small $n_{\pi}/n_{\sigma }$ gives for both cases
\begin{equation}
\label{expansion}
\frac{\Delta_\sigma^0}{T_c} \approx \pi e^{-\gamma}
\left( 1 - \frac{n_{\pi}}{n_{\sigma} }
\left(1+\Lambda\, \ln |\rho_0| \right) \ln |\rho_0 | \rho_0^2 
\right).
\end{equation}

We now discuss the relation between the gap ratios near $T_c$, 
given by $\rho $,  
and at $T=0$, given by $\rho_0$.
When $\Lambda =0$, we have $\rho=\rho_0$.
For small $|\Lambda |$ or small $\rho_0/n $ we obtain from Eq.~(\ref{rho}) the approximate relation
\begin{eqnarray}
\rho \approx \rho_0+\Lambda \rho_0\,\ln |\rho_0| \,.
\label{ratios}
\end{eqnarray}
For $|\rho_0 | < 1$ it follows 
that when the Coulomb repulsion dominates in the second channel 
($\Lambda <0$), the magnitude of the gap ratio at $T_c $ is
increased with respect to that at zero temperature, $|\rho| > |\rho_0 |$;
for effective attraction in the subdominant channel ($\Lambda >0$),
$|\rho |< |\rho_0 |$.

\begin{figure}
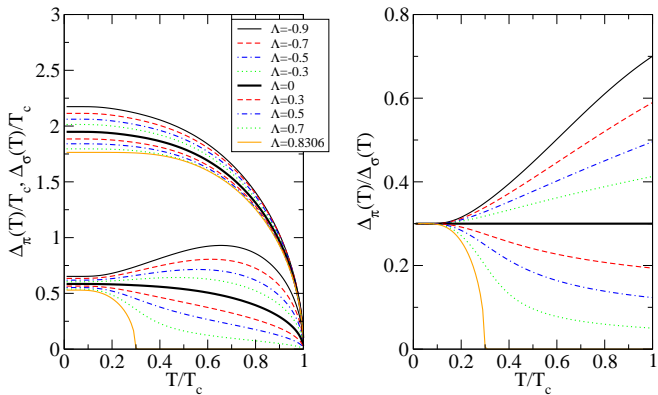

\begin{minipage}{\columnwidth}
\includegraphics[height=0.6\columnwidth]{Hom1b.eps}
\hfill
\includegraphics[height=0.6\columnwidth]{Hom2b.eps}
\end{minipage}
\caption{(Color online) Energy gaps $\Delta_\pi(T)$ and $\Delta_\sigma(T)$
(left-hand panel) and the ratio $\Delta_\pi(T)/\Delta_\sigma(T)$ (right-hand panel) in the 
homogeneous case as a function of temperature $T$ for various values of $\Lambda$,
for fixed $\rho_0=0.3$ and $n=\sqrt{n_{\sigma}/n_{\pi}}=1$.
There is a second transition temperature associated with the $\pi$ band
for $\Lambda\ge\Lambda_c$ (see text).
}
\label{homoorder}
\end{figure}
In Fig.~\ref{homoorder} we show the solutions of the homogeneous
gap equations for both $\sigma $ and $\pi $ bands as a function
of temperature for various values of $\Lambda $, for fixed
$\rho_0=0.3$ and $n=1$.
In the left-hand and right-hand panels of Fig.~\ref{homoorder} the order parameters in the two
bands and the ratio of the two are plotted, respectively, 
as a function of temperature.
As discussed above,
when $\Lambda=0$ (thick lines), the ratio $|\Delta_\pi(T)/\Delta_\sigma(T)|$ stays constant
and equal to $\rho=\rho_0$, while for $\Lambda<0$ ($>0$) 
$\rho$ is larger (smaller) than $\rho_0$.

If the order parameter in the $\pi$ band has a zero slope at $T_c$, i.e.,
$\rho=0$ (see, e.g., the lowest curves in Fig.~\ref{homoorder}), 
then for positive $\Lambda $
there is a second superconducting transition\cite{suhl59}
associated with the $\pi$ band at temperature
\begin{equation}
T_{c\pi }(\rho=0) = T_c e^{-1/\Lambda} \qquad (\Lambda >0).
\end{equation}
In this case the order parameter in
the $\pi $ band stays zero above $T_{c\pi}$;
below $T_{c\pi}$ both order parameters are nonzero, and
the zero-temperature gap ratio is
$\rho_{0} = e^{-1/\Lambda } $
so that the relation $T_{c\pi}/T_c=\rho_0$ holds.
For negative $\Lambda $ there is no second transition in the
$\pi $ band, and superconductivity is purely induced for all temperatures.
For small absolute values of $\rho $, the order parameter in the $\pi $
band is given approximately by
\begin{equation}
\label{smallrho}
\frac{\Delta_\pi(T) }{\Delta_\sigma(T)}
\approx \frac{\rho  }{1-\Lambda 
\left| \ln\frac{T}{T_c} \right| } 
\qquad (|\rho |\ll 1).
\end{equation}
This equation is correct for $T>T_{c\pi}$
as long as $|\Delta_\pi(T) /\Delta_\sigma(T)| $ stays small.
For $\rho \ne 0 $ the order parameter
in the $\pi$ band is nonzero in the entire temperature range
[see Fig.~\ref{homoorder}(a)], being
purely induced for $\Lambda <0$, and for $T>T_{c\pi}$ when $ \Lambda>0$.
It follows from Eq.~(\ref{smallrho}) that for $\Lambda < 0$,
$|\Delta_\pi(T)/\Delta_\sigma(T)|$ is reduced from $|\rho|$ as temperature
decreases,
while for $\Lambda > 0$ it increases from $|\rho |$ for $T_{c\pi}<T<T_c$
[see also Fig.~\ref{homoorder}(b)].

\begin{figure}
\begin{minipage}{\columnwidth}
\includegraphics[width=1.0\columnwidth]{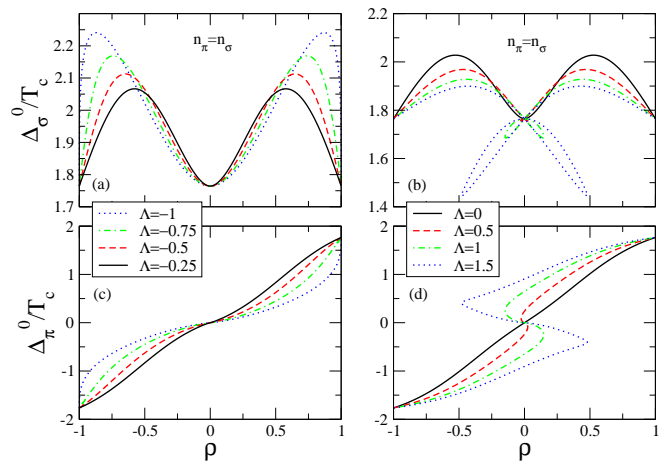}
\end{minipage}
\caption{(Color online) 
The $\Delta^0_\sigma/T_c $ (top) and $\Delta^0_\pi/T_c $
(bottom) as a function of the gap ratio at $T_c$,
$\rho = \lim_{T\to T_c}\Delta_\pi /\Delta_\sigma $, obtained from
Eqs. \eqref{homDs} and \eqref{rho}, for $\Lambda<0$ [(a) and (c)]
and $\Lambda>0$ [(b) and (d)].
While for $-1\le \Lambda \le 0$  there is a unique solution,
for  $\Lambda >0$, for sufficiently small $\rho $  there are three solutions.
This indicates possible first-order transitions as a function of temperature,
if $\Lambda $ for a given $\rho $ is larger than a critical value.
}
\label{Drho}
\end{figure}
For a given $\rho_0$, the critical $\Lambda$, above which
a second transition in the $\pi$ band can be observed, is given by
$\Lambda_c=-1/\ln |\rho_0 |$. In the case of Fig.~\ref{homoorder} this yields
$\Lambda_c\approx 0.8306$.
We find that 
for $\Lambda >\Lambda_c$ there is a $\pi$ phase difference between the
order parameters  in the two bands
for $T>T_{c\pi}$ (the gap ratio
becomes negative), and for $T<T_{c\pi}$ a more complicated picture
arises, with three solutions that compete, and possible 
first-order phase transitions. 
This is demonstrated in Fig.~\ref{Drho}, in which
the zero-temperature bulk gap to $T_c$ ratio in the $\sigma$ (top) and $\pi$ (bottom) band
is plotted as a function of the gap ratio at $T_c$, $\rho = 
\lim_{T\to T_c}\Delta_\pi /\Delta_\sigma $, obtained from
the solution of Eqs.~\eqref{homDs} and \eqref{rho}. 
Results are shown for the effective subdominant coupling $\Lambda $ ranging from repulsive (left) to
attractive (right). 
For $-1\le \Lambda \le 0$ there is a unique solution for the bulk gaps at zero temperature 
for a given $\rho$ (left-hand panels).  However, as can be seen in the right-hand panels,
when $\Lambda>0$, there are three possible solutions to the gap equations for sufficiently small $\rho$.
As these zero-temperature solutions are associated with the same single 
solution near $T_c$,
this indicates that first-order phase transitions can occur as temperature is varied.
This behaviour is not the topic of this paper, however, and we
assume for the remainder of the paper that $\Lambda<\Lambda_c$ for
a given $\rho_0$ (or $\rho_0>\rho_c=e^{-1/\Lambda }$ for a given $\Lambda $).

\subsection{Choice of material parameters}
\label{comp3}
For MgB$_2$, if estimated from the {\it ab initio} values of the 
electron-phonon coupling strengths\cite{liu01,golubov02,choi02} and the 
Coulomb pseudopotentials,\cite{golubov02,moon04} $\Lambda$ varies
depending on which calculations are used,
while some calculations yield very small $\Lambda$.
But using Eq.~(\ref{rho}), one can fix $\Lambda$ from the observed
values of $\rho_0$ and $\rho$ -- although for the latter, some data points have 
relatively large error bars.  From the experiments of 
Refs.~\onlinecite{iavarone02}, \onlinecite{szabo01,gonnelli02,holanova04}, and
\onlinecite{tsuda05},
we find roughly $-0.3 < \Lambda < 0.3$.  Among these, the measurement of 
Ref.~\onlinecite{iavarone02} has a data point at the highest temperature, 
$T\simeq 0.96 T_c$, and $\Lambda$ estimated using this point is close to zero.  
Therefore, we assume that for MgB$_2$, $\Lambda$
is small and can be either positive or negative.

The strength of induced superconductivity in the $\pi$ band is specified
in our model
by $\Lambda$ and $\rho_0$, the latter of which is about 0.3 in MgB$_2$.  
Results are presented mainly for $\rho_0=0.1,0.3,0.5$.
The $n_{\pi}/n_{\sigma}$ in MgB$_2$ has been determined experimentally to be
$n_{\pi}/n_{\sigma}\approx 1-1.2$.\cite{eskildsen02,junod01,sologubenko02}
We show results for $n_{\pi}/n_{\sigma}=1$ and 2, and discuss the effects
of the $\pi$-band density of states.

Finally, there is
another material parameter in our model, that is the ratio of
the coherence lengths in the two bands, $\xi_\pi/\xi_\sigma$.
Note in this regard that in defining the $\pi $-band coherence length 
$\xi_\pi=\sqrt{D/2\pi T_c}$,
we have used the energy scale $T_c$ rather than 
$\Delta_{\pi }$.
This is motivated by the fact that, as discussed above, 
superconductivity in the $\pi $ band 
is assumed to be mostly induced by the $\sigma $ band. Consequently, 
for the variation of the $\pi $-band order parameter determined 
by the self-consistency equation, Eq.~(\ref{gapeq}), 
the energy scale $T_c$ is relevant.
However, the {\it quasiparticle motion} in the
$\pi$ band is governed by the Eilenberger transport equation supplemented 
with an appropriate impurity self-energy $\Sigma_{i,\pi}$ of the order of 
$ 1/\tau_\pi $, where $\tau_\pi $ is the $\pi $-band
quasiparticle lifetime due to elastic intraband impurity scattering.
To determine under which conditions the
diffusive approximation can be used in the $\pi$ band, 
the relative size of the $\pi$-band gap, $\Delta_\pi $, and 
the $\pi$ band impurity self-energy,
$\Sigma_{i,\pi}$, matters.
Thus, the condition for the $\pi$ band to be diffusive so that the Eilenberger
equation supplemented with $\Sigma_{i,\pi }$ reduces to 
the Usadel equation is 
\begin{equation}
1/\tau_{\pi } > \Delta_\pi.
\end{equation}
In practise it turns out that the diffusive approximation applies
quite well already when $1/\tau_{\pi } $ is only larger,
not much larger, than $\Delta_\pi$.
Using the above definitions for $\xi_\sigma$ and $\xi_\pi$ and
with the diffusion constant $D={1\over 3}\tau_{\pi} v_{F\pi}^2$,
this yields
\begin{equation}
{\xi_\pi\over\xi_\sigma}< {v_{F\pi}\over v_{F\sigma}}\sqrt{2\pi T_c\over 3\Delta_\pi}.
\label{xicond}
\end{equation}
Strong electron-phonon coupling in the $\sigma$ band renormalizes 
the Fermi velocity as $v_{F\sigma}/Z$,\cite{nicol05} where 
$Z$ is the renormalization factor and is about 2 in the zero-frequency
limit.\cite{dolgov05}
Taking the {\it ab initio} Fermi velocities, 
$v_{F\pi}/v_{F\sigma}=5.8/4.4$,\cite{brinkman02} 
as the unrenormalized values, and using $\Delta_\pi/T_c\approx 0.6$
as found in several experiments, we estimate $\xi_\pi/\xi_\sigma < 5$ for MgB$_2$.
As discussed in Sec.~\ref{total_ldos}, an estimate from the experiment
of Ref.~\onlinecite{eskildsen02} yields a value between 1 and 3 for MgB$_2$.
To keep the discussion general,
the effects of the coherence length ratio on the vortex core structure
are illustrated for $\xi_\pi/\xi_\sigma=1,2,3,5$.

Throughout this work, cylindrical symmetry is assumed for the two-dimensional 
$\sigma$ band, and an isotropic diffusion tensor for the three-dimensional
$\pi $ band.
Up to this point, the energy gaps (homogeneous order parameters) 
in the two bands have been denoted as
$\Delta_\pi$ and $\Delta_\sigma$.
In the following, $\Delta_\pi$ and $\Delta_\sigma$ represent the spatially varying
order parameters.

\section{Results}
\label{results}
\subsection{Order parameter}
\label{order}

\subsubsection{Effects of the Coulomb interaction in the diffusive band}
\label{interaction}

\begin{figure}
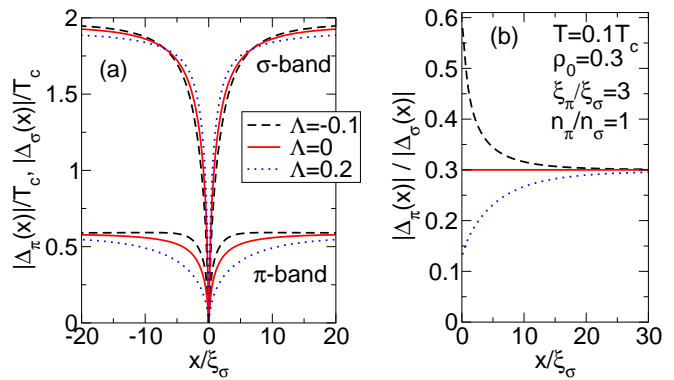

\begin{minipage}{\columnwidth}
\includegraphics[height=0.57\columnwidth]{Order1.eps}
\hfill
\includegraphics[height=0.57\columnwidth]{Order1a.eps}
\end{minipage}
\caption{(Color online) (a) Order parameters $|\Delta_\pi(x)|$ and $|\Delta_\sigma(x)|$
and (b) the ratio $|\Delta_\pi(x)/\Delta_\sigma(x)|$
as a function of coordinate $x$ along the path $y=0$ for various values of 
$\Lambda$,
for $\xi_\pi/\xi_\sigma=3$, $\rho_0=0.3$, $T/T_c=0.1$, $n=1$.
The Coulomb repulsion (which reduces $\Lambda $) renormalizes the recovery lengths of the order parameters
in the two bands differently.
}
\label{order1}
\end{figure}

Negative $\Lambda $ means that the effective 
Coulomb interaction dominates over the effective 
pairing interaction in the second pairing channel.
In this case, Coulomb interactions are responsible for the fact that
the $\pi$ band cannot maintain superconductivity on its own, and it 
superconducts
only due to its proximity to the superconducting $\sigma $ band.
We first examine the effects of the Coulomb interaction
on the order parameters.
We consider a vortex that extends in $z$ direction, with the vortex centre 
situated at $x=y=0$ for each $z$.
In Fig.~\ref{order1}, we show the order parameter magnitudes as a 
function of coordinate $x$ along the path $y=0$ for 
three values of $\Lambda $, appropriate for attractive, repulsive
and no effective interaction in the second pairing channel.
Figure~\ref{order1}(a) 
shows $|\Delta_\pi(x)|$ and $|\Delta_\sigma(x)|$, and
Fig.~\ref{order1}(b) their ratio.
The Coulomb repulsion renormalizes the recovery lengths of the order 
parameters
(the characteristic length scale over which the order parameter 
recovers to the bulk value)
in both bands. 
Although superconductivity in the $\pi$ band is mostly induced by that 
in
the $\sigma $ band, 
it is clear from Fig.~\ref{order1} that
the length scales 
in the two bands can differ considerably if $\Lambda \ne 0$.
For $\Lambda=0$ there is only one length scale and hence the same 
recovery
length in both bands.  
When $\Lambda<0$ ($>0$),
$|\Delta_\pi(x)|$ is enhanced (suppressed) around the vortex core, 
resulting in a shorter (longer)
recovery length compared to the case with $\Lambda=0$. On the other 
hand,
the sign of $\Lambda$ has opposite effects on $|\Delta_\sigma(x)|$, 
although these effects are
relatively small for the parameter set for Fig.~\ref{order1}.
In fact, $\Delta_\sigma$ (its length scale and magnitude) can be 
affected 
by the $\pi$-band Coulomb interaction significantly
for larger $\rho_0$ and $\xi_\pi/\xi_\sigma$.\cite{tanaka06}

As demonstrated in Fig.~\ref{gaptc}, 
the bulk gap over $T_c$ ratios in both bands are enhanced by stronger Coulomb 
repulsion in the diffusive band, an effect also
seen in Fig.~\ref{order1}(a).  
As shown in Eq.~(\ref{lambda0}),
this renormalization of the magnitude of the bulk gap
has the same sign for both bands, and is very 
different from
the renormalization in the core region seen in Fig.~\ref{order1}(a),
that has opposite sign for the
two order parameters. The latter effect can be seen most dramatically when plotting
the ratio $|\Delta_\pi(x)/\Delta_\sigma(x)|$, as we do 
in Fig.~\ref{order1}(b). For $\Lambda \ne 0$ the ratio increases by up 
to
a factor of 2 for $\Lambda=-0.1$, and decreases by about the same factor
for $\Lambda=0.2$. When $\Lambda=0$, 
$|\Delta_\pi(x)/\Delta_\sigma(x)|$ stays constant and equal to 
$\rho_0=\rho$.
This results from the fact that in this case $\lambda^{(1)}=0$ in 
Eq.~\eqref{gapk}, leading to $\Delta^{(1)}(x)=0$ so that 
Eq.~\eqref{gap0only}
holds for all $x$.

\subsubsection{Effects of the density of states in the diffusive band}
\label{dos}

\begin{figure}
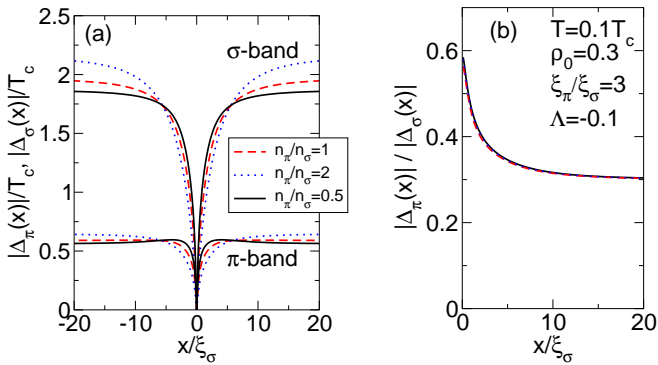

\begin{minipage}{\columnwidth}
\includegraphics[height=0.55\columnwidth]{Order2.eps}
\hfill
\includegraphics[height=0.55\columnwidth]{Order2a.eps}
\end{minipage}
\caption{(Color online) (a) Order parameters $|\Delta_\pi(x)|$ and $|\Delta_\sigma(x)|$
and (b) the ratio $|\Delta_\pi(x)/\Delta_\sigma(x)|$
as a function of coordinate $x$ along the path $y=0$ for various density of states ratios
$n_{\pi}/n_{\sigma}$, and fixed
$\xi_\pi/\xi_\sigma=3$, $\rho_0=0.3$, $T/T_c=0.1$, $\Lambda=-0.1$.
The $\pi$-band density of states renormalizes the recovery lengths of the order parameters,
and the bulk magnitudes of the order parameters are larger for larger $n_{\pi}/n_{\sigma}$.
}
\label{order2}
\end{figure}

Figure~\ref{order2} demonstrates the effects of the Fermi-surface 
density of states in the diffusive band on the order parameters
for a typical set of material parameters and for $T/T_c=0.1$.
In Fig.~\ref{order2}(a) we show
the order parameter magnitudes and in (b) their ratio.
It can be seen in Fig.~\ref{order2}(a) that 
the bulk gap over $T_c$ ratios increase with increasing density of states ratio $n_{\pi }/n_{\sigma }$,
a behavior consistent with Fig.~\ref{gaptc}(b) and with
Eqs.~\eqref{homDs} and \eqref{expansion}.
From Fig.~\ref{order2}(a) we see that 
a larger $n_{\pi }/n_{\sigma }$ also enlarges the core area in both bands.
At higher temperature (not shown), $n_{\pi }/n_{\sigma }$ has less effect on the order parameters.
The effects of the $\pi$-band Coulomb interaction described
in the last section manifest 
more drastically in both bands for larger $n_{\pi }/n_{\sigma }$.
When comparing the shapes of the order parameter profiles 
in the $\sigma $ band and the $\pi $ band in Fig.~\ref{order2}(a),
there is a qualitative difference between the cases $n_{\pi }/n_{\sigma }=0.5$
and $n_{\pi }/n_{\sigma }=2$. Nevertheless, Fig.~\ref{order2}(b) shows that
despite the different shapes of the order parameter profiles in the
two bands for different values of $n_{\pi }/n_{\sigma }$,
the spatial variation of the {\it ratio}  $|\Delta_\pi(x)/\Delta_\sigma(x)|$ 
is almost independent of  $n_{\pi }/n_{\sigma }$. This is for $\Lambda \ne 0 $ a 
nontrivial and rather interesting finding.

\subsubsection{Effects of the coherence length and gap ratios}
\label{coherence_gap}

\begin{figure}
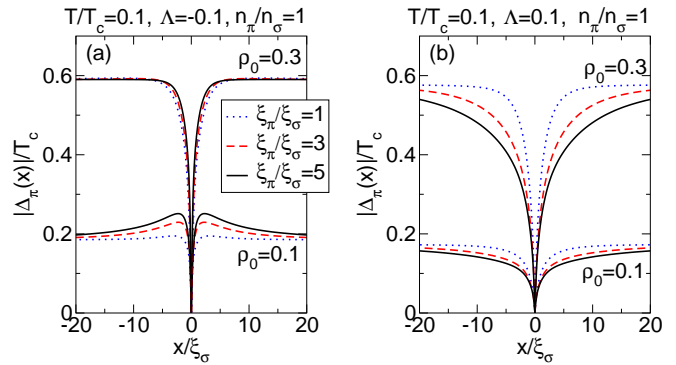

\begin{minipage}{\columnwidth}
\includegraphics[height=0.55\columnwidth]{Order3.eps}
\hfill
\includegraphics[height=0.55\columnwidth]{Order3b.eps}
\end{minipage}
\caption{(Color online) 
(a) Order parameter in the $\pi$ band $|\Delta_\pi(x)|$
as a function of coordinate $x$ along the path $y=0$ for various values of 
$\rho_0$ and $\xi_\pi/\xi_\sigma$, for $T/T_c=0.1$ and  $n_{\pi}/n_{\sigma}=1$;
for (a)  $\Lambda=-0.1$ and (b)  $\Lambda=0.1$.
For $\Lambda>0$, $|\Delta_\pi(x)|$ is depleted around the vortex core 
for larger $\xi_\pi/\xi_\sigma$, and this effect is larger for larger $\rho_0$.
When the Coulomb repulsion dominates ($\Lambda<0$), for small $\rho_0$ (e.g., 0.1),
$|\Delta_\pi(x)|$ is enhanced near the vortex centre for larger $\xi_\pi/\xi_\sigma$.
}
\label{order3b}
\end{figure}
In Fig.~\ref{order3b} we illustrate how the order parameter profile in the $\pi$ band
changes as $\rho_0$ and $\xi_\pi/\xi_\sigma$ are varied. We compare the
two cases of (a) $\Lambda=-0.1$ and (b) $\Lambda=0.1$, leaving
$T/T_c=0.1$ and $n_\pi/n_\sigma=1$ fixed.
As can be seen in Fig.~\ref{order3b}(b),
for $\Lambda>0$, $|\Delta_\pi(x)|$ is depleted in the core area as $\xi_\pi/\xi_\sigma$ increases,
and this effect is larger for larger $\rho_0$. On the other hand,
when the Coulomb repulsion is dominant in the $\pi$ band ($\Lambda <0$),
as shown in Fig.~\ref{order3b}(a),
the profile of $|\Delta_\pi(x)|$  changes in a peculiar way as $\xi_\pi/\xi_\sigma$ varies.
For relatively small $\rho_0$ [e.g., 0.1 in Fig.~\ref{order3b}(a)],
$|\Delta_\pi(x)|$ is enhanced around the vortex core for larger $\xi_\pi/\xi_\sigma$,
while for  $\rho_0=0.3$
$|\Delta_\pi(x)|$ hardly changes for different values of  $\xi_\pi/\xi_\sigma$.
For larger $\rho_0(>0.3)$ (not shown), however, the core region becomes larger with increasing 
$\xi_\pi/\xi_\sigma$ -- in a similar way as for $\Lambda>0$, though not so drastically.
This lets us conclude that for the case $\rho_0\approx 0.3 $ in
Fig.~\ref{order3b}(a) a cancellation
between two opposite tendencies is at work, that renders $|\Delta_\pi(x)|$
seemingly insensitive to the parameter $\xi_\pi/\xi_\sigma$.
As to $|\Delta_\sigma(x)|$, similarly as $|\Delta_\pi(x)|$ in Fig.~\ref{order3b}(b),
the core area is enlarged for larger $\xi_\pi/\xi_\sigma$, more noticeably 
for larger $\rho_0$, 
and for both positive and negative $\Lambda$ [see Fig.~\ref{order4}(b) below].

\begin{figure}
\begin{minipage}{\columnwidth}
\includegraphics[width=1.0\columnwidth]{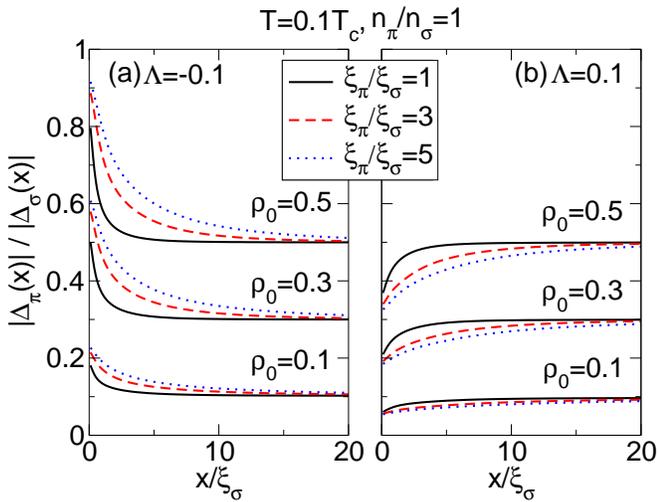}
\end{minipage}
\caption{(Color online) (a) The ratio $|\Delta_\pi(x)/\Delta_\sigma(x)|$
as a function of coordinate $x$ along the path $y=0$ for various values of 
$\rho_0$ and $\xi_\pi/\xi_\sigma$, for $T/T_c=0.1$ and  $n_{\pi}/n_{\sigma}=1$;
for (a)  $\Lambda=-0.1$ and (b)  $\Lambda=0.1$.
For negative (positive) $\Lambda$, the ratio is larger (smaller) than $\rho_0$
around the vortex core, and the area of the enhancement (depletion)
is larger for larger $\xi_\pi/\xi_\sigma$.
As $\rho_0$ increases, the ratio changes more drastically as $x$ approaches zero.
}
\label{order3a}
\end{figure}

As discussed above, even though superconductivity is induced in the $\pi$ band due to coupling
with the $\sigma$ band, the characteristic length scale of the order parameter
can be quite different in the two bands.
This is demonstrated in Fig.~\ref{order3a} for various sets of $\rho_0$ and $\xi_\pi/\xi_\sigma$.
Here the order parameter ratio $|\Delta_\pi(x)/\Delta_\sigma(x)|$
is plotted as a function of $x$.
In agreement with our discussion of Fig.~\ref{order1}(b), this ratio is in the vortex core
larger than $\rho_0$ for negative, and smaller than $\rho_0$ 
for positive $\Lambda$.
When $\Lambda<0$, despite the peculiar change of the $|\Delta_\pi(x)|$ profile as a function of
$\xi_\pi/\xi_\sigma$ depending on the value of $\rho_0$, 
$|\Delta_\pi(x)/\Delta_\sigma(x)|$ is always larger than $\rho_0$
in the core area, as can be seen in Fig.~\ref{order3a}(a).
The main conclusion of
Fig.~\ref{order3a} is that the enhancement for $\Lambda<0$ and depletion for $\Lambda >0$
of $|\Delta_\pi(x)/\Delta_\sigma(x)|$ 
is increasing with $\xi_\pi/\xi_\sigma$ and $\rho_0$.

\begin{figure}
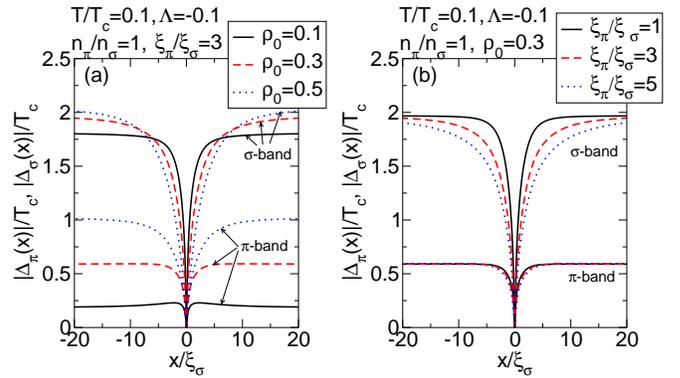

\begin{minipage}{\columnwidth}
\includegraphics[height=0.57\columnwidth]{Order4.eps}
\hfill
\includegraphics[height=0.57\columnwidth]{Order4b.eps}
\end{minipage}
\caption{(Color online) Order parameters $|\Delta_\pi(x)|$ and $|\Delta_\sigma(x)|$
as a function of coordinate $x$ along the path $y=0$ for $T/T_c=0.1$, $\Lambda=-0.1$,
$n_{\pi}/n_{\sigma}=1$; for (a) various values of $\rho_0$ for 
$\xi_\pi/\xi_\sigma=3$ and (b) different values of $\xi_\pi/\xi_\sigma$ for $\rho_0=0.3$.
For a fixed $\xi_\pi/\xi_\sigma$, as $\rho_0$ increases, 
the core area is enhanced in both bands.
For a fixed $\rho_0$, for larger $\xi_\pi/\xi_\sigma$ the core area is enlarged in the $\sigma$ band.
}
\label{order4}
\end{figure}
In Fig.~\ref{order4} we compare the influence of the material parameters $\rho_0 $ and $\xi_\pi/\xi_\sigma$
on the order parameter profiles in the vortex core.
In Fig.~\ref{order4}(a) we
vary $\rho_0$ for fixed $\xi_\pi/\xi_\sigma=3$, and in (b) we vary $\xi_\pi/\xi_\sigma$ for fixed $\rho_0=0.3$.
As shown in Fig.~\ref{gaptc}, in general
the bulk gap to $T_c$ ratio in the $\sigma$ band changes in a nonmonotonous way as a function of $\rho_0$.
In the case of the parameter set for Fig.~\ref{order4}(a), however, the bulk $|\Delta_\sigma|/T_c$
increases with increasing $\rho_0$ for the values of $\rho_0$ shown (for a fixed $\xi_\pi/\xi_\sigma$).
At the same time,
as the coupling of the two bands becomes stronger,
the recovery length of $|\Delta_\sigma|$ increases [Fig.~\ref{order4}(a)]
and the length scales in the two bands are more different [see also Fig.~\ref{order3a}(a)].
For a fixed $\rho_0$ the $\sigma$-band order parameter has a larger
core area as $\xi_\pi/\xi_\sigma$ increases [Fig.~\ref{order4}(b)].
The $|\Delta_\sigma|$ profile changes in a similar way with varying $\xi_\pi/\xi_\sigma$ when $\Lambda>0$.
Also seen in Fig.~\ref{order4}(b) is that the insensitivity of
the $|\Delta_\pi |$ profile due to the cancellation effect for
$\Lambda<0 $ discussed above,
combined with the enhanced depletion of $|\Delta_\sigma|$, yields
a seemingly paradoxical picture, namely that an increase in $\xi_\pi$ (for fixed
$\xi_\sigma $) results in
a larger core area in the $\sigma $ band instead of the $\pi $ band.

\subsubsection{Temperature dependence}
\label{temperature}

\begin{figure}
\begin{minipage}{\columnwidth}
\includegraphics[width=1.0\columnwidth]{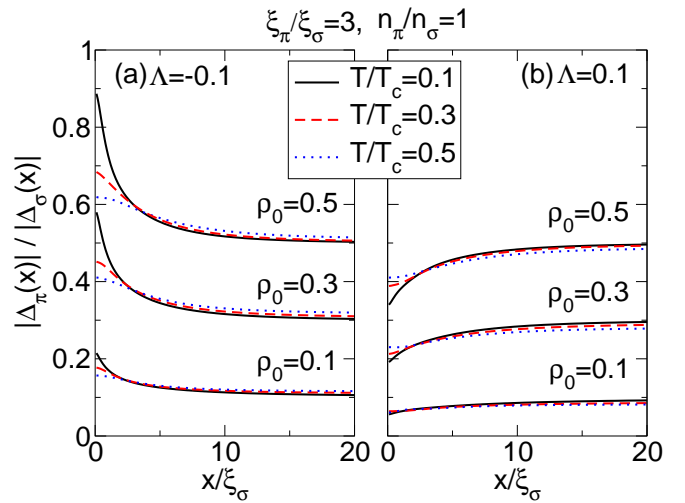}
\end{minipage}
\caption{(Color online) The ratio $|\Delta_\pi(x)/\Delta_\sigma(x)|$
as a function of position $x$ along the path $y=0$ for $\xi_\pi/\xi_\sigma=3$ and 
$n_{\pi}/n_{\sigma}=1$, for various values of $\rho_0$ and  $T/T_c$;
for (a) $\Lambda=-0.1$ and (b) $\Lambda=0.1$.
The difference between the length scales in the two bands and its temperature dependence
are enhanced as $\rho_0$ increases, especially for $\Lambda<0$.
As temperature is raised, the difference between the recovery lengths of the
two order parameters becomes smaller.
}
\label{order4a}
\end{figure}

\begin{figure}
\begin{minipage}{\columnwidth}
\includegraphics[width=1.0\columnwidth]{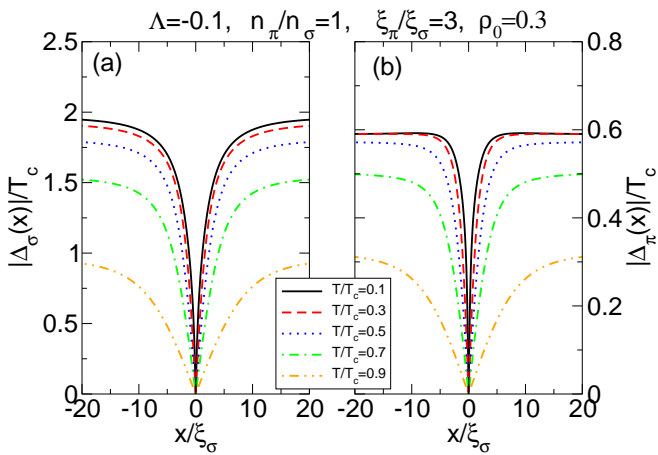}
\end{minipage}
\caption{(Color online) Order parameters (a) $|\Delta_\sigma(x)|$ and (b)  $|\Delta_\pi(x)|$ 
as a function of $x$ along the path $y=0$ for
$\rho_0=0.3$, $\xi_\pi/\xi_\sigma=3$, $n_{\pi}/n_{\sigma}=1$, $\Lambda=-0.1$.
The `weak' $\pi$ band makes the temperature dependence of the $|\Delta_\sigma|$ profile
more drastic than in the single-band case.
}
\label{order5}
\end{figure}

The ratio $|\Delta_\pi(x)/\Delta_\sigma(x)|$ as a function of coordinate $x$
is plotted in Fig.~\ref{order4a} 
for various values of $\rho_0$ and $T/T_c$, for (a) $\Lambda=-0.1$ and (b) $\Lambda=0.1$.
The enhancement (depletion) and its temperature dependence 
of this ratio in the core area are larger for larger $\rho_0$ for negative (positive) $\Lambda$.
This effect is considerable when the Coulomb repulsion dominates in the $\pi$ band 
[Fig.~\ref{order4a}(a)]. 
As temperature becomes higher, however, the difference between the recovery lengths of the
two order parameters is reduced, as can be seen in Fig.~\ref{order4a} (see also Fig.~\ref{order5}).

Figure~\ref{order5} shows the temperature dependence of
(a) $|\Delta_\sigma(x)|$ and (b)  $|\Delta_\pi(x)|$
for a typical set of material parameters.
By coupling to the weak $\pi$ band, the suppression of $|\Delta_\sigma|$
and the widening of its core region with increasing temperature are more drastic
compared with the single-band case.  The $\pi$-band order parameter is less affected
by temperature for low enough $T/T_c$.  For higher temperature, however, the profile of
the order parameters in the two bands are similar.

The temperature dependence of the core area of the order parameter can be characterised by
the vortex `core size' defined by\cite{kato01}
\begin{equation}
\xi_{c}^{-1} = {\partial \Delta(r=0)\over \partial r}
{1\over \Delta(r=\infty)}\,,
\label{coresize}
\end{equation}
where $r$ is the radial coordinate measured from the vortex centre.
In a clean single-band superconductor,
the order parameter exhibits the Kramer-Pesch (KP) effect,\cite{kramer74} i.e., shrinkage of the 
vortex core as $T$ is lowered, approaching zero in the zeo-temperature limit
\cite{kato01,ichioka96,hayashi05,gumann05}.
In an $s$-wave superconductor, however, the core size as defined above saturates as
temperature approaches zero, when there are nonmagnetic impurities.\cite{hayashi05,gumann05}
This can be understood by the broadening of the vortex core bound states
that carry the supercurrent around the vortex centre. This broadening
removes the singular behaviour in the spatial variation of the order
parameter in the vortex centre, and the vortex core shrinking ceases
when $k_BT$ becomes smaller than the energy width of the zero-energy bound
states situated at the vortex centre (in quasiclassical approximation
one can neglect the small splitting
of these bound states due to the Caroli-deGennes minigap).

Recently, our model of coupled clean and dirty bands\cite{tanaka06} has been applied to study the
KP effect in a two-band superconductor, and it has been shown that,
by coupling to the clean band, the KP effect is induced in the dirty band.\cite{gumann05}
The calculation in Ref.~\onlinecite{gumann05} corresponds to positive $\Lambda$ in
our formulation.
\begin{figure}
\begin{minipage}{\columnwidth}
\includegraphics[width=1.0\columnwidth]{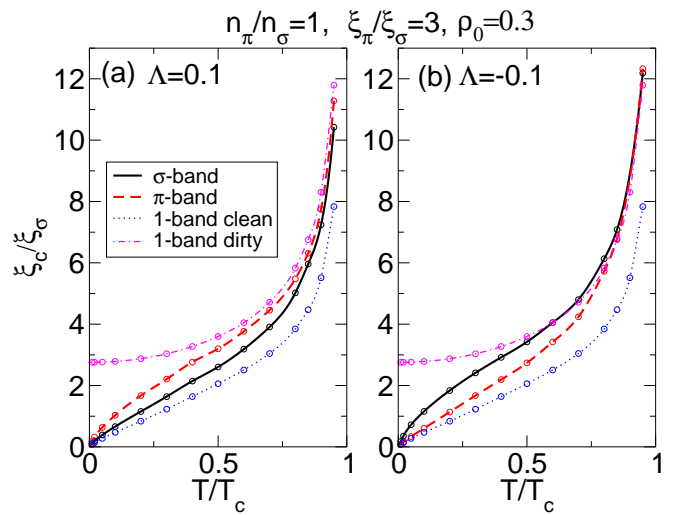}
\end{minipage}
\caption{(Color online) 
The core size $\xi_{c}$ defined by Eq.~(\ref{coresize}) in the two bands for
$\rho_0=0.3$, $\xi_\pi/\xi_\sigma=3$, and $n_{\pi}/n_{\sigma}=1$,
for (a)  $\Lambda=0.1$ and (b) $\Lambda=-0.1$.
The Kramer-Pesch effect is induced in the $\pi$ band, although absent 
in a single diffusive band (dashed-dotted curves). 
As seen from (a),
the core radius $\xi_{c}$ is larger in the $\pi $ band
than that in the $\sigma$ band when $\Lambda>0$. However, the opposite is
true for $\Lambda<0$, as shown in (b).
}
\label{order6}
\end{figure}

In Fig.~\ref{order6} we plot $\xi_{c}/\xi_{\sigma}$ in the $\sigma$ and $\pi$ bands
as a function of temperature for 
$\rho_0=0.3$, $\xi_\pi/\xi_\sigma=3$, and $n_{\pi}/n_{\sigma}=1$, for (a) $\Lambda=0.1$
and (b) $\Lambda=-0.1$.
Points are data obtained by self-consistent calculation, and
curves are guides to the eye.
Single-band results are also shown for clean (dotted curves) and dirty 
(dashed-dotted curves) superconductors,
which exhibit the core shrinkage and saturation as $T\rightarrow 0$, 
respectively.
In the two-band case, the core size in the dirty $\pi $ band (dashed curves)
tends to zero in the zero-temperature limit, as the core size in the
clean $\sigma $ band (full curves) does.
For $\Lambda>0$, shown in Fig.~\ref{order6}(a), the core size in the 
$\pi $ band is always larger than the core size in the $\sigma $ band.
There is the KP effect in the $\pi$ band
also when the Coulomb repulsion is dominant [see Fig.~\ref{order6} (b)].  
However, here
the core size defined by Eq.~(\ref{coresize}) is always smaller 
in the $\pi$ band than in the $\sigma$ band.
Surprisingly, with dominating Coulomb interactions,
the KP effect is better developed rather in the diffusive band than in the ballistic band.
Only at very low temperature
the core size extrapolates to zero in the $\sigma $ band, whereas this
happens in the $\pi $ band already below $T\sim 0.5 T_c$.
We suggest that the KP effect can be used to study the strength of Coulomb 
interactions in the effective pair interaction matrix.
Finally, when $\Lambda=0$ (not shown), 
the order parameters in the two bands have the same length scale and the same
$\xi_{c}$.

\subsection{Spectral properties of diffusive band}
\label{ldos_dirty}

\begin{figure}
\begin{minipage}{\columnwidth}
\includegraphics[width=1.00\columnwidth]{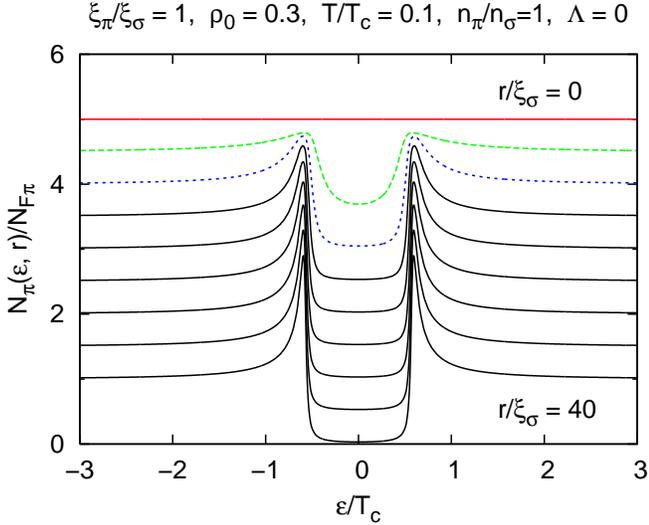}
\end{minipage}
\caption{(Color online) The $\pi$-band LDOS $N_\pi(\epsilon,r)$ as a function of
energy $\epsilon$ at various distances $r$ from the vortex centre
($r/\xi_\sigma$ from 0 to 40 with an increment of 5).
The parameter values are $\xi_\pi/\xi_\sigma=1$,  $\rho_0=0.3$, $T/T_c=0.1$, $n_\pi/n_\sigma=1$, 
and $\Lambda=0$.
The $N_\pi(\epsilon,r)$ shows no sign of localized states in the vortex core,
consistent with the experiment on MgB$_2$ (Ref.~\cite{eskildsen02}).
}
\label{usadel1}
\end{figure}

In Fig.~\ref{usadel1} we show an example for the LDOS $N_\pi(\epsilon,r)$ as obtained from
Eq.~\eqref{LDOS} in the diffusive $\pi$ band 
as a function of energy $\epsilon$ at various distances $r$ from the vortex centre.
At the vortex centre, $N_\pi(\epsilon,r=0)$ as a function of $\epsilon$ is completely flat,
showing no sign of localized states. This is consistent with 
the experiment on MgB$_2$,\cite{eskildsen02} which probed the $\pi$-band LDOS by 
tunneling along the $c$ axis.
Far away from the vortex core, the bulk BCS density of states is
recovered.  For a fixed $\rho_0$, the BCS density of states is recovered at larger distances
for larger $\xi_\pi/\xi_\sigma$; while for a fixed  $\xi_\pi/\xi_\sigma$, 
the LDOS recovers to the bulk density of states further away for smaller $\rho_0$.

\begin{figure}
\begin{minipage}{\columnwidth}
\includegraphics[width=1.00\columnwidth]{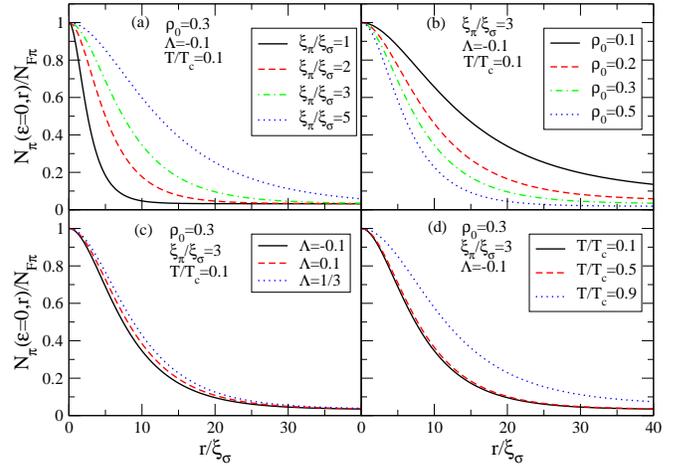}
\end{minipage}
\caption{(Color online) 
The zero-bias LDOS $N_\pi(\epsilon=0,r)$ as a function of distance $r$ from the vortex centre
for $n_\pi/n_\sigma=1$; (a) $\rho_0=0.3$, $T/T_c=0.1$, $\Lambda=-0.1$ for various values of 
$\xi_\pi/\xi_\sigma$; (b) $\xi_\pi/\xi_\sigma=3$, $T/T_c=0.1$, $\Lambda=-0.1$ for various values of 
$\rho_0$; (c) $\rho_0=0.3$, $\xi_\pi/\xi_\sigma=3$, $T/T_c=0.1$ for various values of $\Lambda$;
(d) $\rho_0=0.3$, $\xi_\pi/\xi_\sigma=3$, $\Lambda=-0.1$ for various temperatures.
By comparing with the experimental data (Ref.~\onlinecite{eskildsen02}) we estimate $\xi_\pi/\xi_\sigma\sim 2$
for  MgB$_2$, taking $\rho_0=0.3$.
}
\label{usadel2}
\end{figure}

Experimentally the vortex `core size' obtained from tunneling spectroscopy
can be defined as a measure of
the decay of the zero-bias LDOS as one moves away from the vortex centre.
Note that this definition of the vortex core size is very different from that
defined in Eq.~(\ref{coresize}) and shown in Fig.~\ref{order6}: the two
quantities in fact differ by a large amount as we discuss in the following.
Figure~\ref{usadel2} shows the zero-bias LDOS in the $\pi$ band $N_\pi(\epsilon=0,r)$
as a function of $r$ 
for (a) varying $\xi_\pi/\xi_\sigma $, (b) varying $\rho_0 $, (c) varying $\Lambda $, and
(d) varying $T/T_c$; for the combination of the remaining parameters
$n_\pi/n_\sigma=1$, $\xi_\pi/\xi_\sigma=3$, $\rho_0=0.3$, $T/T_c=0.1$, $\Lambda=-0.1$.
For a fixed $\rho_0$, the core size defined as a decay length of the zero-bias LDOS increases
substantially with increasing $\xi_\pi/\xi_\sigma$, as shown in Fig.~\ref{usadel2}(a).
On the other hand, for a fixed $\xi_\pi/\xi_\sigma$, the core size becomes considerably 
larger for weaker coupling to the $\sigma$ band, i.e., smaller $\rho_0$, as
presented in Fig.~\ref{usadel2}(b).
It can be seen in Fig.~\ref{usadel2}(c) that the LDOS profile in the $\pi$ band
is barely changed for different values
of $\Lambda$. The Fermi-surface density of states $n_\pi/n_\sigma$ also has little effect on
the $\pi$-band LDOS (not shown).
As illustrated in Fig.~\ref{usadel2}(d), the LDOS is hardly affected by 
temperature over a rather large temperature range;
however, the decay length increases notably for temperatures close to $T_c$.
Fig.~\ref{usadel2} is a central result of our calculations, as it 
can be used directly as reference for comparison with
scanning tunneling experiments, in order to determine either of the material parameters 
$\xi_\pi/\xi_\sigma $ or $\rho_0 $. In particular, the parameter
$\xi_\pi/\xi_\sigma $ is otherwise difficult to access experimentally.

The decay length of the LDOS can be very different in the two bands.
The length scale of the zero-bias LDOS in the $\sigma$ band is $\xi_\sigma$
and thus shorter than that in the $\pi$ band for small enough $\rho_0$  or
large  enough $\xi_\pi/\xi_\sigma $.
The existence of two coherence length scales in MgB$_2$ has been suggested;
\cite{eskildsen02,serventi04,kohen05} i.e., a `core size' ($\sim$50 nm)
much larger than that expected from $H_{c2}$ ($\sim$10 nm).
Theoretically, two apparent length scales in the LDOS have been found also
in the case of two clean bands\cite{nakai02} and two dirty bands.\cite{koshelev03,knote}
The difference in the decay lengths comes from the fact that, although 
superconductivity is induced in the $\pi $ band, quasiparticle motion
is governed by the effective coherence length resulting from
the Usadel equation (\ref{usadeleq}), 
$\sqrt{D/2\Delta_{\pi }}$ (that is longer than $\xi_\pi=\sqrt{D/2\pi T_c}$).

Our model of a clean and a dirty band is suitable for describing MgB$_2$,
and the values of $\rho_0$, $n_\pi/n_\sigma$ and $\Lambda$ for Fig.~\ref{usadel2}(a)
are appropriate for modelling MgB$_2$.
Using the {\it ab initio} value\cite{brinkman02} of $v_{F\sigma}$ and including
the renormalization factor as mentioned above, $\xi_\sigma\simeq 6.8$nm for MgB$_2$
($T_c=39$ K). By comparing Fig.~\ref{usadel2}(a) with the 
zero-bias LDOS measured in the experiment of Ref.\onlinecite{eskildsen02}
[Fig.~3(a) in Ref.~\onlinecite{eskildsen02}], for this parameter set,
we find $\xi_\pi/\xi_\sigma$ is about 2 for MgB$_2$.
As discussed in detail in the next section,
there can be bound states near the gap edge in the `strong' ballistic $\sigma$ band,
which arise from coupling to the `weak' diffusive $\pi$ band.
Such bound states exist for parameter values relevant for MgB$_2$, e.g., 
for $\xi_\pi/\xi_\sigma=2$, $\rho_0=0.3$, and 
for various temperatures, $n_\pi/n_\sigma$, and $\Lambda$ [see, e.g., Fig.~\ref{edge1}].

\subsection{Spectral properties of ballistic band}

\subsubsection{Total LDOS}
\label{total_ldos}

\begin{figure}
\begin{minipage}{\columnwidth}
\includegraphics[width=1.00\columnwidth]{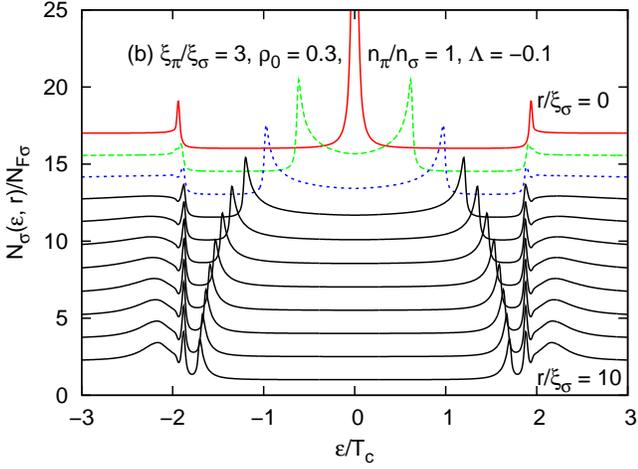}
\end{minipage}
\caption{(Color online)
The $\sigma$-band LDOS $N_\sigma(\epsilon,r)$ as a function of
energy $\epsilon$ at various distances $r$ from the vortex centre
($r/\xi_\sigma$ from 0 to 10 with an increment of 1).
The parameter values are $\xi_\pi/\xi_\sigma=3$, $\rho_0=0.3$, $T/T_c=0.1$, $n_\pi/n_\sigma=1$, 
and $\Lambda=-0.1$.
Coupling to a diffusive band results in bound states at the gap edge in the clean band, in addition to the
well-known Caroli-de Gennes-Matricon bound states.
}
\label{ldose1}
\end{figure}

\begin{figure}
\begin{minipage}{\columnwidth}
\includegraphics[width=1.00\columnwidth]{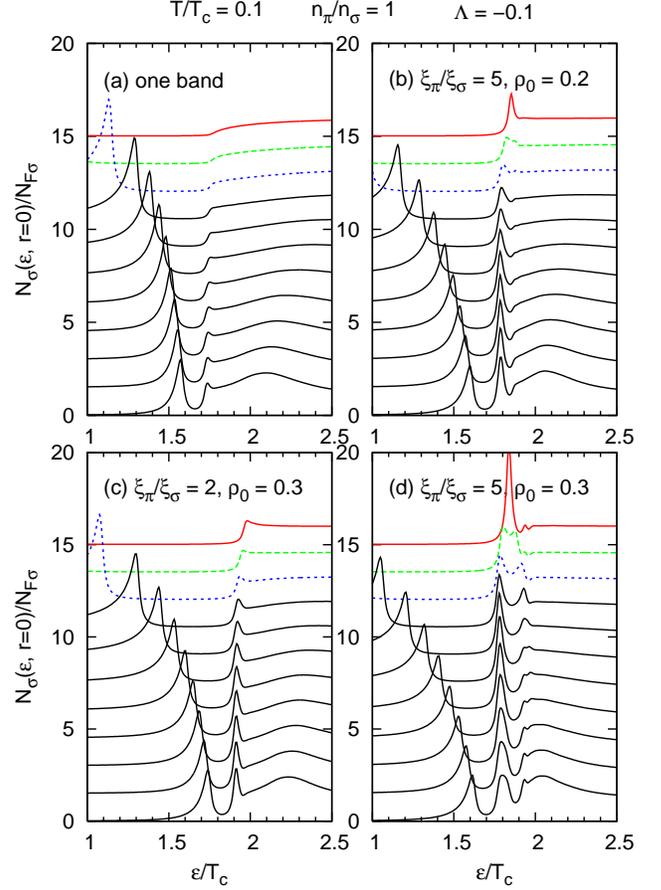}
\end{minipage}
\caption{(Color online) 
The  $N_\sigma(\epsilon,r)$ as a function of
energy $\epsilon$ at various distances $r$ from the vortex centre
($r/\xi_\sigma$ from 0 to 10 with an increment of 1) for $T/T_c=0.1$; (a) the single-band case,
(b) $\xi_\pi/\xi_\sigma=5$, $\rho_0=0.2$, (c) $\xi_\pi/\xi_\sigma=2$, $\rho_0=0.3$,
and (d) $\xi_\pi/\xi_\sigma=5$, $\rho_0=0.3$.
In the two-band cases, $n_\pi/n_\sigma=1$ and $\Lambda=-0.1$.
The gap-edge bound states, which are absent in the single-band case, are enhanced
(and the number of branches increases) for larger  $\xi_\pi/\xi_\sigma$ and $\rho_0$.
}
\label{edge1}
\end{figure}
We turn now to the spectral properties of the vortex core originating from the clean $\sigma $ band.
We plot in Fig.~\ref{ldose1} the total (angle-averaged) LDOS
$N_{\sigma}(\epsilon,r)=\langle N_{\sigma}(\epsilon,{\bf p}_{F\sigma}, r) \rangle_{{\bf p}_{F\sigma}}$
in the ballistic $\sigma$ band
as a function of energy $\epsilon$ at various distances $r$ from the vortex centre.
The spectra show the well-known Caroli-de Gennes-Matricon (CdGM) bound-state bands\cite{caroli64,bardeen69} at low
energies. In the single-band case [see Fig.~\ref{edge1}(a)], at the gap edge 
the spectrum is suppressed and there is neither coherence peak nor bound state
in the vortex centre spectrum.\cite{waxman96,matthias0,matthias3}
The new feature found in our model is the additional bound states at the gap edge, as can be seen
clearly in Fig.~\ref{ldose1}, which arise from coupling to a diffusive band.
While the CdGM bound states disperse strongly as a function of position, the extra bound states
stay near the gap edge as $r$ is varied.
The self-consistency of the order parameters is essential for studying the 
presence or absence of these gap-edge bound states.

In Fig.~\ref{edge1} the total LDOS near the gap edge at various distances $r$ is shown
at temperature $T/T_c=0.1$, for (a) the single-band case,
(b) $\xi_\pi/\xi_\sigma=5$, $\rho_0=0.2$, (c) $\xi_\pi/\xi_\sigma=2$, $\rho_0=0.3$,
and (d) $\xi_\pi/\xi_\sigma=5$, $\rho_0=0.3$.
It can be seen in Fig.~\ref{edge1}(a) that in the single-band case,
there is no localized state near the gap edge at $r=0$, while
a small peak develops at the gap edge at large $r$.
This is a bound state arising solely from the phase winding around the vortex.
In Fig.~\ref{edge1}(b), additional bound states exist at the gap edge
both in the core region and far away from the vortex centre.
For a fixed $\xi_\pi/\xi_\sigma$, with increasing $\rho_0$, gap-edge bound states are enhanced and more
bound states can be created  [Figs.~\ref{edge1}(b) and \ref{edge1}(d)].
In the case of Fig.~\ref{edge1}(d), there are two bound states close to
the gap edge in the $r=0$ spectrum.
For nonzero $r$, all bound states broaden into bands 
due to dispersion of the bound states with momentum direction,
exhibiting the typical $1/\sqrt{\epsilon}$-like behaviour at the band edges.
This can be seen clearly in Fig.~\ref{ldose1} for the main bound state
crossing zero energy at $r=0$.  The additional gap-edge bound states also show
this behaviour, as can be seen, e.g., in Fig.~\ref{edge1}(d) 
for the main gap-edge bound state at small $r$.
For a fixed $\rho_0$, (more) bound states can exist for larger $\xi_\pi/\xi_\sigma$
[Figs.~\ref{edge1}(c) and ~\ref{edge1}(d)]. In the case of Fig.~\ref{edge1}(c),
gap-edge bound states exist at large $r$, but not in the core region.
However, in this case a coherence peak can be observed at the gap edge at small $r$.

\subsubsection{Gap-edge bound states}
\label{gapedge_bs}

\begin{figure}
\begin{minipage}{\columnwidth}
\includegraphics[width=1.00\columnwidth]{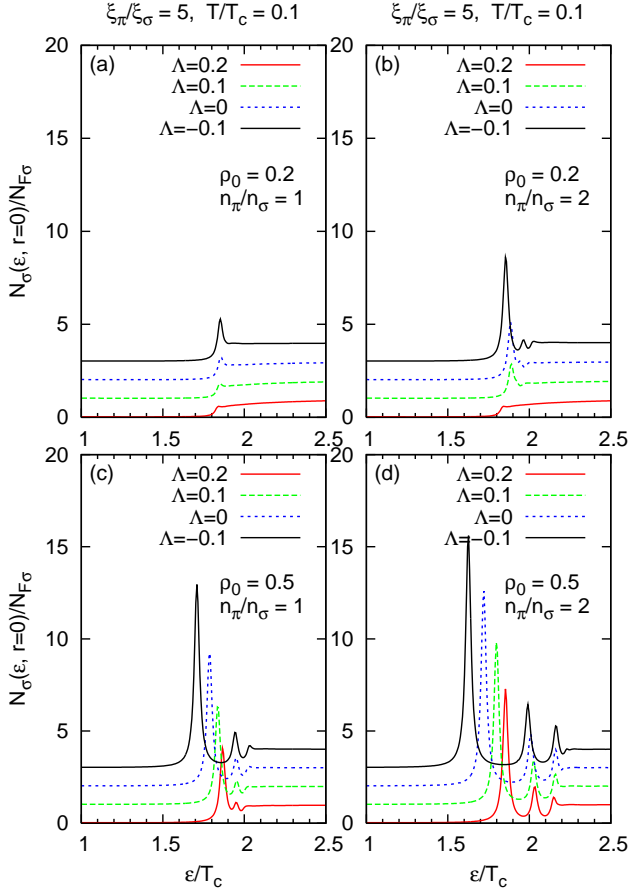}
\end{minipage}
\caption{(Color online) The total LDOS in the $\sigma$-band at the vortex centre,
$N_{\sigma}(\epsilon,r=0)$, for energy $\epsilon$ 
close to the energy gap, for $\xi_\pi/\xi_\sigma = 5$ and $T/T_c = 0.1$
for different values of $\Lambda$; for $\rho_0=0.2$ (a,b) and 0.5 (c,d)
and for $n_\pi/n_\sigma=1$ (a,c) and 2 (b,d).
Bound states exist near the gap edge for $\Lambda$ smaller than a certain value.
}
\label{gapedge1}
\end{figure}
We now study the detailed development of bound states near the gap edge in
the $\sigma $ band in terms of the spectrum at the vortex centre.
We start with the sensitivity of the gap-edge bound state spectrum to
$\Lambda $ and to $n_\pi/n_\sigma $.
In Fig.~\ref{gapedge1} we show the $\sigma$-band total LDOS $N_{\sigma}(\epsilon,r=0)$
for energy $\epsilon$ 
close to the energy gap at temperature $T=0.1T_c$. 
To emphasize the changes we show here examples for a relatively
large coherence length ratio $\xi_\pi/\xi_\sigma = 5$.
The panels (a)-(d) correspond to various fixed parameter combinations, and 
the curves in each panel are for different values of $\Lambda$.
We show results for $\rho_0=0.2$ (a,b) and 0.5 (c,d), and for
and for $n_\pi/n_\sigma=1$ (a,c) and 2 (b,d).
The Coulomb interaction in the $\pi$ band also affects the gap-edge bound states.
It can be seen in Fig.~\ref{gapedge1}(a) that, for the given set of 
parameters, gap-edge bound states exist for $\Lambda=-0.1$, but not for 
$\Lambda=0.2$.  
An increase in $n_\pi/n_\sigma$ 
results in an increase in the energy gap, and (more) gap-edge bound states due to widening of
the core area [Fig.~\ref{gapedge1}(b)].
Increasing $\rho_0$ has similar effects. It can be seen in Figs.~\ref{gapedge1}(c,d) that
the lowest branch shifts considerably to lower energy, while the energy gap becomes larger, 
as $\Lambda$ is reduced.

\begin{figure}
\begin{minipage}{\columnwidth}
\includegraphics[width=1.00\columnwidth]{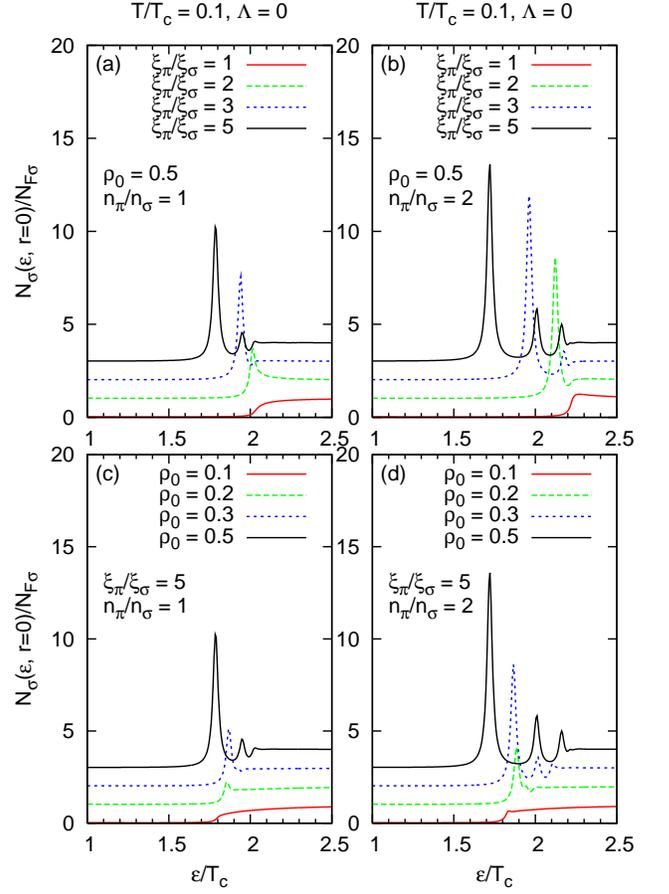}
\end{minipage}
\caption{(Color online) Same as Fig.~\ref{gapedge1}, but for fixed $\Lambda=0$,
and for (a) various values of $\xi_\pi/\xi_\sigma$ for $\rho_0=0.5$ and
(b) different values of $\rho_0$ for $\xi_\pi/\xi_\sigma=5$.
Gap-edge bound states develop as $\rho_0$ or $\xi_\pi/\xi_\sigma$ is increased.
}
\label{gapedge2}
\end{figure}
Next we discuss the dependence on the coherence length ratio $\xi_\pi/\xi_\sigma$ and on
the zero temperature gap ratio $\rho_0$.
There is a threshold for both parameters, above which gap-edge bound states appear.
We illustrate this in Fig.~\ref{gapedge2} for $T/T_c = 0.1$ and $\Lambda=0$.
In Figs.~\ref{gapedge2}(a,b) $\xi_\pi/\xi_\sigma$ is changed for fixed $\rho_0=0.5$.
Results are for (a) $n_\pi/n_\sigma=1$ and 2 (b).  
For this gap ratio, gap-edge bound states exist for $\xi_\pi/\xi_\sigma > 1$ for both ratios of the
density of states shown. As $\xi_\pi/\xi_\sigma$ increases, the energy gap
changes only slightly. On the other hand, the lowest bound-state energies are reduced substantially
and more branches appear. This effect is enhanced for larger $n_\pi/n_\sigma$.
For $\xi_\pi/\xi_\sigma=5$ and $\rho_0=0.5$, there are three branches for $n_\pi/n_\sigma=2$.
Figures ~\ref{gapedge2}(c,d) show the LDOS for various values of $\rho_0$ for fixed
 $\xi_\pi/\xi_\sigma=5$ for (c) $n_\pi/n_\sigma=1$ and 2 (d). 
For this parameter set, for $\rho_0=0.1$ there is no gap-edge bound state for 
$n_\pi/n_\sigma\leq 2$.  As $\rho_0$ increases, the energy gap
is enhanced and more branches appear at the gap edge, and this
effect is more significant for larger  $n_\pi/n_\sigma$. 

\begin{figure}
\begin{minipage}{\columnwidth}
\includegraphics[width=1.00\columnwidth]{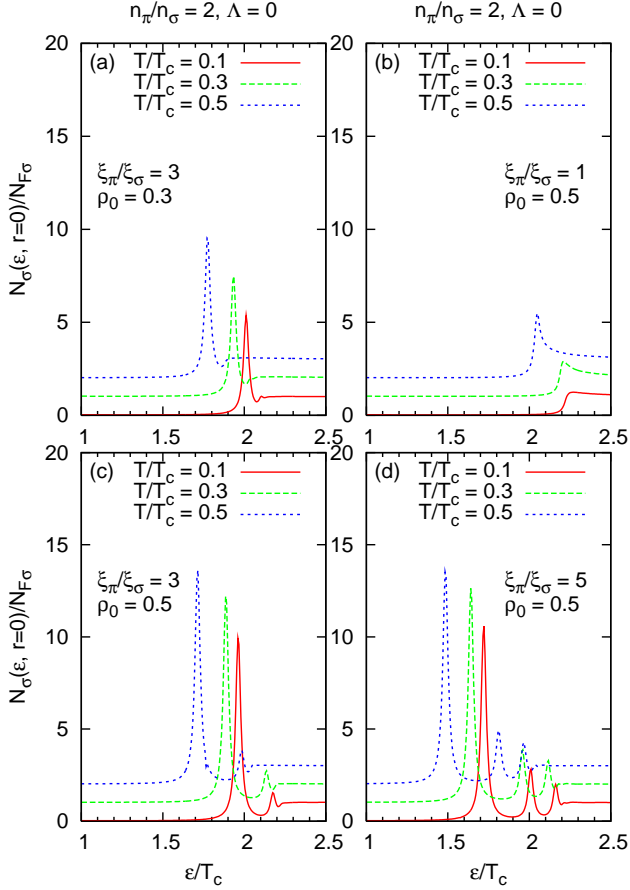}
\end{minipage}
\caption{(Color online) Same as Fig.~\ref{gapedge1}, but for
$n_\pi/n_\sigma=2$ and $\Lambda=0$, for four sets of other parameters (see text)
and various temperatures.
The energy gap is reduced significantly as temperature increases and the energies
of the gap-edge bound states are shifted together with the energy gap.
}
\label{gapedge3}
\end{figure}
Increasing temperature $T$ also enhances the features of gap-edge bound states.
Figure~\ref{gapedge3} shows the LDOS for $n_\pi/n_\sigma=2$ and $\Lambda=0$
for various $T$.
In (a) and (c) we compare for $\xi_\pi/\xi_\sigma=3$ the cases
$\rho_0=0.3$ and $\rho_0=0.5$, and in (b) and (d) we fix $\rho_0=0.5$ and
compare the cases $\xi_\pi/\xi_\sigma=1$ and $\xi_\pi/\xi_\sigma=5$.
As $T$ increases, bound states at the gap edge shift lower together with the
energy gap and increase in strength. 
The change from $T/T_c=0.1$ to 0.5 in Fig.~\ref{gapedge3} is substantial for
$\rho_0=0.5$ and $\xi_\pi/\xi_\sigma=5$.
The results in Fig.~\ref{gapedge1}-\ref{gapedge3}
illustrate that the gap edge bound states are a robust feature of our model.

\subsubsection{Angle-resolved LDOS}
\label{angle_ldos}

\begin{figure}
\begin{minipage}{\columnwidth}
\includegraphics[width=1.00\columnwidth]{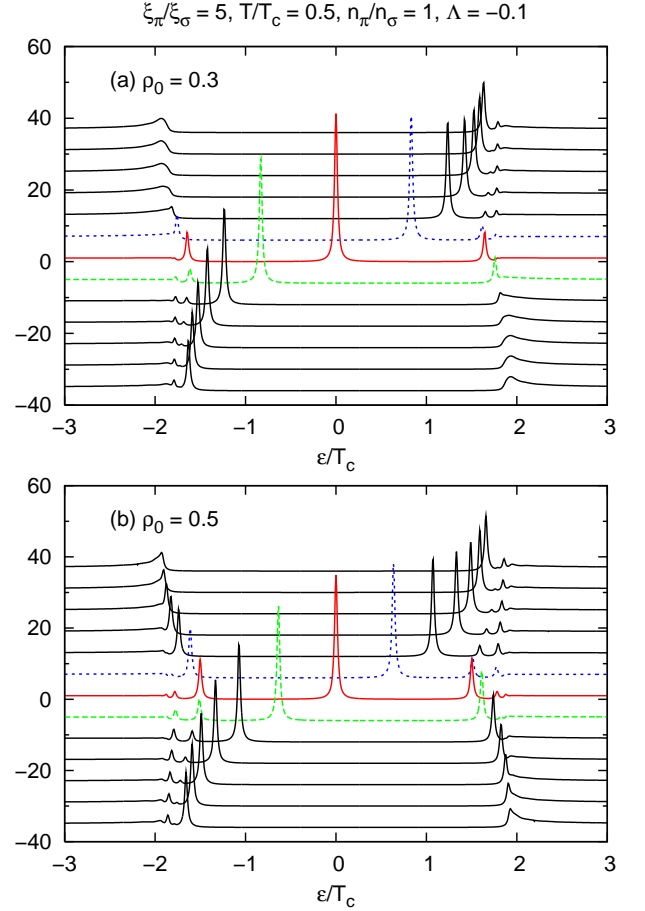}
\end{minipage}
\caption{(Color online) The angle-resolved LDOS $N_\sigma(\epsilon,p_y=0,y)$ 
as a function of energy $\epsilon$, for quasiparticles moving in the $x$ direction 
at various positions along the $y$ axis
(from $y=-18\xi_\sigma$ to $y=18\xi_\sigma$ with an increment of 3$\xi_\sigma$).
The parameter values are $\xi_\pi/\xi_\sigma=5$,  $T/T_c=0.5$,  $n_\pi/n_\sigma=1$,
$\Lambda=-0.1$, with (a) $\rho_0=0.3$ and (b) $\rho_0=0.5$.
An increase in $\rho_0$ results in more gap-edge bound states and enhancement
of their dispersion.
}
\label{angle1}
\end{figure}

\begin{figure}
\begin{minipage}{\columnwidth}
\includegraphics[width=0.70\columnwidth]{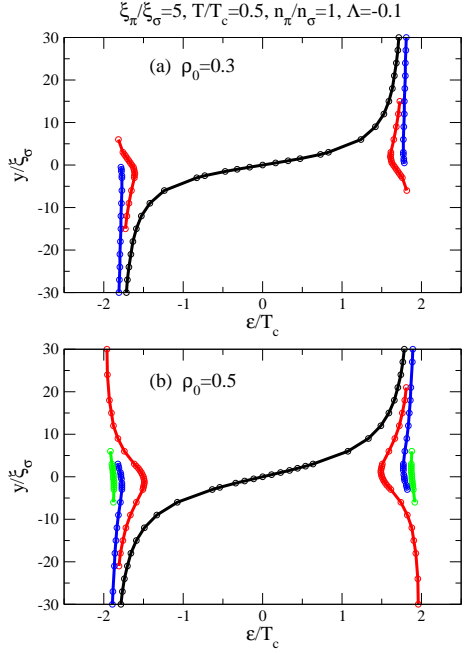}
\end{minipage}
\caption{(Color online)
The energy of the bound states as a function of $y$ obtained from Fig.~\ref{angle1}.
There are two branches of gap-edge bound states for $\rho_0=0.3$.
For $\rho_0=0.5$ there is a third branch in the vortex core: the lowest branch is
extended over large distances and its dispersion is enhanced considerably 
compared with the case for $\rho_0=0.3$.
}
\label{peak1}
\end{figure}

The bound-state spectrum can be discussed most clearly in terms of the
angle-resolved LDOS spectra, obtained from Eq.~\eqref{LDOS}.  In Fig.~\ref{angle1}, 
the angle-resolved LDOS $N_\sigma(\epsilon,p_y=0,y)$ for 
quasiparticles moving with impact parameter $y$ in the $p_x$ direction is shown as a function of
$\epsilon$ at various positions $y$ ($x=0$).
The two panels are for (a) $\rho_0=0.3$ and (b) $\rho_0=0.5$.
The position of the bound states as a function of impact parameter $y$ is plotted in Fig.~\ref{peak1}.
The CdGM bound-state branch disperses with angular momentum,
crossing the chemical potential in the vortex centre.  In Fig.~\ref{angle1}(a)
additional bound states near the gap edge can be seen, which have only weak dispersion.
A close inspection reveals that there are two branches;
for $\epsilon>0$, the lower branch for roughly $y>-6$ and the higher one for $y>0$,
and similarly for negative $y$ for $\epsilon<0$. 
The higher branch hardly disperses with $y$ 
[see Fig.~\ref{peak1}(a)].
As coupling with the $\pi$ band becomes stronger, 
the energy gap increases.  At the same time, the energies of the bound states 
are lowered (for relatively small angular momenta),
and more bound states appear near the gap edge,
due to enlargement of the core area [Fig.~\ref{angle1}(b)] 
[see also Fig.~\ref{peak1}(b)].
For $\rho_0=0.5$, there is a third bound-state branch at the gap edge
for small $|y|$, and the lowest branch extends over large distances, with
enhanced dispersion compared to that for $\rho_0=0.3$.
For a fixed $\rho_0$, increasing $\xi_\pi/\xi_\sigma$ results in a larger core area
and has similar effects as increasing $\rho_0$.

\subsection{Current density}
\label{current}

\begin{figure}
\begin{minipage}{\columnwidth}
\includegraphics[width=1.00\columnwidth]{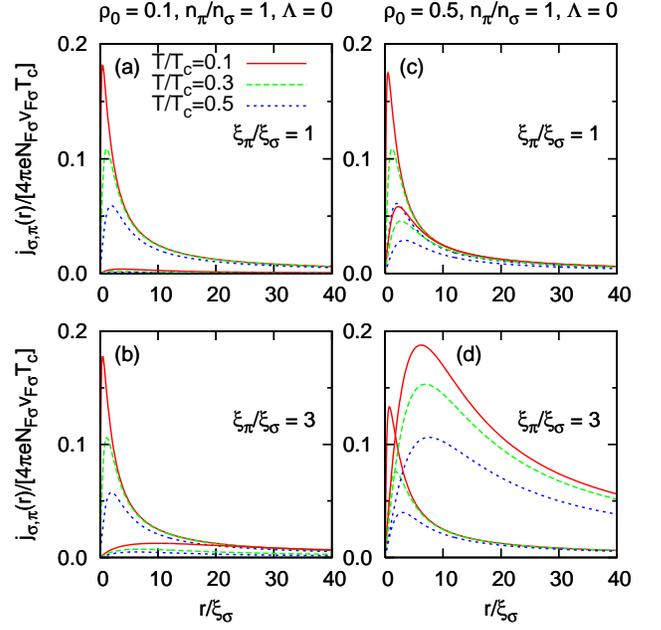}
\end{minipage}
\caption{(Color online) Magnitude of the supercurrent density in the two bands,
$j_\pi$ and $j_\sigma$, as a function of distance $r$ from the vortex centre,
for $n_\pi/n_\sigma=1$ and $\Lambda=0$, for various temperatures.
In the left-hand panels, we have $\rho_0=0.1$, for (a) $\xi_\pi/\xi_\sigma =1$
and (b) 3.  The right-hand panels have $\rho_0=0.5$, 
for (c)  $\xi_\pi/\xi_\sigma =1$ and (d) 3.
The $\pi$-band contribution can be substantial, or even dominating, and have strong
temperature dependence far outside the vortex core
for large enough $\rho_0$ and $\xi_\pi/\xi_\sigma$.
}
\label{current1}
\end{figure}

For a clean single-band superconductor,
supercurrents around a vortex are strongly coupled to the Andreev spectrum
discussed in the preceding section, and are inside the core area carried mainly 
by these bound states.\cite{bardeen69,waxman96}
In the following we present results for the 
current density calculated with Eqs.~\eqref{CD} for our two-band model.
In Fig.~\ref{current1} we show contributions from the two bands to the
current density around the vortex separately, 
for parameters $n_\pi/n_\sigma=1$ and $\Lambda=0$, and for various temperatures.
In the left-hand panels, we have $\rho_0=0.1$, and (a) $\xi_\pi/\xi_\sigma =1$
and (b) 3. For this weak coupling between the two bands,
the current contribution from the $\pi$ band is negligible for $\xi_\pi/\xi_\sigma =1$.
For $\xi_\pi/\xi_\sigma =3$, though still small, at low temperature (i.e., $T/T_c=0.1$)
the $\pi$-band contribution in the bulk (outside the vortex core, but still
well within the penetration depth distance from the core)
is almost the same as that in the $\sigma$ band.
As $T\to 0$, the peak of $j_\sigma$ approaches the vortex centre -- this is the KP effect
manifest in the supercurrent.
The current density arising from the induced superconductivity
in the $\pi$ band is also enhanced by decreasing temperature, but
the maximum does not approach the vortex centre [see Fig.~\ref{current1}(b)].
In Ref.~\onlinecite{tanaka06},
the same trend of the $\pi$-band current density as a function of temperature
was presented for parameter values
appropriate for MgB$_2$. 

As $\rho_0$ increases, 
the $\pi$-band contribution to the current density becomes considerable.
In the right-hand panels of Fig.~\ref{current1}, the current density contributions are
plotted for $\rho_0=0.5$, for (c)  $\xi_\pi/\xi_\sigma =1$ and (d) 3.
Interestingly, as the coupling between the two bands is increased, the
$\pi$ band starts exhibiting the KP effect in the current density.
The maximum shifts towards the vortex centre as temperature is decreased,
although the `core size' as defined by the position of the current maximum remains
larger than that in the $\sigma$ band [see Figs.~\ref{current1}(c) and (d)].

As can be seen in Fig.~\ref{current1}(d),
for relatively large $\rho_0$ and $\xi_\pi/\xi_\sigma(>1)$, the $\pi$-band contribution 
to the current density becomes substantial.  Especially away from the vortex core, where the 
contribution of the $\sigma $ band is reduced, the $\pi$-band contribution can be
dominant.  Another important observation is that the current density
in the $\pi$ band has strong temperature dependence also far outside the core.
On the contrary,
the $\sigma$-band current density is temperature-dependent only 
in the core area.  Far away from the vortex core, 
we can use in Eq.~(\ref{CD})
the analytical solutions to the Eilenberger and Usadel equations 
for a quasihomogeneous system, with constant magnitude and constant phase gradient
of the order parameters.
We thus obtain the current-density magnitudes
for a large distance $r$ from the vortex centre
(but small compared to the London penetration depth) as
\begin{eqnarray}
j_{\sigma }({r})&\approx& eN_{F\sigma}  v_{F\sigma }^2{1\over 2r}\,,\\
j_{\pi}({r}) &\approx& eN_{F\pi} \pi D |\Delta_{\pi }|{1\over r}\,.
\label{bulk_currents}
\end{eqnarray}
Clearly, the temperature dependence of $j_\pi$ is dominated by that of
$|\Delta_\pi|$.
This bulk behaviour of the current density in the two bands also explains the
dominance of the $\pi$-band contribution in the bulk for relatively large $\rho_0$ 
and $\xi_\pi/\xi_\sigma$.  For large $r$, the current-density ratio approaches
\begin{eqnarray}
\lim_{r\gg \xi_{\pi,\sigma} }\frac{j_{\pi}({r})}{ j_{\sigma}({r})}
=\frac{N_{F\pi}2\pi D |\Delta_{\pi }|}{N_{F\sigma}v_{F\sigma }^2}
= \frac{n_{\pi}}{n_{\sigma}}\frac{|\Delta_{\pi }|}{T_c} \left(\frac{\xi_\pi }{\xi_\sigma }\right)^2.
\label{ratio}
\end{eqnarray}
Thus the $\pi$-band contribution to the current density 
dominates when 
$(\xi_\pi /\xi_\sigma )^2 > (n_{\sigma}/n_{\pi})(|\Delta_\pi |/T_c)^{-1}$.
This is relevant for MgB$_2$, for which this condition roughly reads as 
$\xi_\pi /\xi_\sigma>1$.

\section{Conclusions}
\label{conclusion}

In conclusion, we have studied a model recently introduced by us\cite{tanaka06} 
for describing a multiband superconductor
with a ballistic and a diffusive band
in terms of coupled Eilenberger and Usadel equations. 
Both equations were solved directly and numerically until self-consistency of
the order parameters was achieved.
We have studied the effects of induced superconductivity
and impurities in the weak diffusive ($\pi$) band on 
the order parameter, the current density, and the spectral properties 
of the strong ballistic ($\sigma$) band.
A unique feature found in our model is the existence of additional bound states
at the gap edge in the ballistic band, which are absent when there is no coupling 
with the diffusive band.

Although the two bands are coupled by the pairing interaction,
the order parameters in the two bands can have very different length scales.
When the Coulomb interaction dominates in the $\pi$ band, the order parameter in
the $\sigma$ band has a longer recovery length than that in the $\pi$ band.
In this case, the $\pi$ band exhibits the Kramer-Pesch effect with
a `core size' [defined by Eq.~(\ref{coresize})] smaller than that in the $\sigma$ band.
Furthermore, when coupled to the diffusive band, the order parameter in 
the ballistic $\sigma$ band is suppressed by temperature more strongly than in the single-band
case.

The zero-bias LDOS in the two bands can have very different decay lengths, the one in
the $\pi$ band being larger than that in the $\sigma$ band.
As the induced superconductivity in the $\pi$ band becomes stronger, 
the core area widens in the $\sigma$ band and (more) gap-edge bound states appear.
Moreover, an increase in the Fermi-surface density of states in the $\pi$ band results in an
increase in the energy gap and the number of gap-edge bound states in the $\sigma$ band.
Increasing the Coulomb repulsion in the $\pi$ band and temperature also enlarges
the core area in the $\sigma$ band, which results in additional bound states
at the gap edge.
The gap-edge bound states have only weak dispersion.  
It is thus expected that these bound states are affected only weakly by impurity
scattering within the $\sigma$ band.\cite{matthias3}
Results incorporating impurities in the strong $\sigma$ band confirm this
statement and will be presented in a future paper.\cite{tanaka2}

The supercurrent density is dominated in the vortex core by the $\sigma$-band
contribution, and outside the core the $\pi$-band contribution can be 
substantial, or even dominating over the $\sigma$-band current density.
The current density in the $\sigma$ band shows the
Kramer-Pesch effect in the core area.  On the other hand, the $\pi$ band current
density has strong temperature dependence also outside the core. It exhibits the
Kramer-Pesch effect when the coupling between the two bands is relatively strong.
Stronger Coulomb repulsion in the $\pi$ band enhances the current density in the
$\pi$ band, while it has little effect on the $\sigma$-band current density.

Our model is suitable for describing MgB$_2$.  For parameter values appropriate
for MgB$_2$, we have found the above intriguing features in the LDOS and the
current density, including additional bound states near the gap edge in the $\sigma$ band.
The gap-edge bound states should be affected only weakly by the strong electron-phonon
interaction, as the energy of relevant phonons is much higher (above 60-70 meV) 
(Refs.~\onlinecite{geerk05,naidyuk03,yanson04}).
Our predictions on the spectral properties of the $\sigma$ band
can be tested by tunneling electrons onto the $ab$ plane.\cite{iavarone02,iavarone05}

\section{Acknowledgements}
Two of the authors (K.T. and M.E.) would like to thank
Morten R. Eskildsen, Boldizs\'ar Jank\'o, Juha Kopu, and Gerd Sch\"on for
valuable discussions and contributions. Two of the authors
(K.T. and D.F.A.) are grateful to G. W. Crabtree and W. K. Kwok for hospitality
at the Argonne National Laboratory, where a part of this work was performed,
and to M. Iavarone, W. K. Kwok, P. C. Moca, and H. Schmidt for discussions.
The research was supported by the 
NSERC of Canada, the Canada Foundation for Innovation,
the Deutsche Forschungsgemeinschaft within the CFN,
the U.S. DOE, Basic Energy Sciences (Contract No. W-7405-ENG-36),
the NSF (Grant No. DMR-0381665),
and the American Chemical Society Petroleum Research Fund.


\begin{thebibliography} {999}
\bibitem{knote}
In the work of Ref.~\onlinecite{koshelev03}, the diffusion constant in the $\pi$ band was
assumed to be larger than that in the $\sigma$ band, for the $\pi$ band to have a
longer decay length of the zero-bias LDOS.
Experimentally it has been
found that $D_\pi \ll D_\sigma$ in dirty samples (Refs.~\onlinecite{braccini05} and 
\onlinecite{gurevich03}).
\end{thebibliography}
\end{document}